\documentclass[%
 reprint,
superscriptaddress,
 amsmath,amssymb,
 aps,
 prl,
 noeprint
]{revtex4-2}

\usepackage{graphicx}
\usepackage{dcolumn}
\usepackage{bm}
\usepackage{hyperref}
\usepackage[mathlines]{lineno}
\usepackage{braket}
\usepackage{xcolor}
\usepackage{physics}
\usepackage{dsfont}
\usepackage[top=2cm, bottom=2cm, left=1.5cm, right=1.5cm]{geometry}
\usepackage{upgreek}
\hypersetup{
    colorlinks=true,
    citecolor=blue,
    linkcolor=blue,
    filecolor=blue,  
    urlcolor=blue,
    pdftitle={Something from Nothing},
    pdfpagemode=FullScreen,
    }
\usepackage{todonotes}
\setuptodonotes{fancyline,size=\tiny,color=green!30,shadow}

\usepackage{orcidlink}

\newcommand{\kl}{k_{{l}}}

\newcommand{\rmi}{{i}}
\newcommand{\rmd}{{d}}

\newcommand{\ad}{a^{\dagger}}
\newcommand{\bd}{b^{\dagger}}

\begin{document}





\preprint{arXiv:xxxx.xxxxx}

\title{Something from Nothing: \\ Enhanced Laser Cooling of a Mechanical Resonator via Zero-Photon Detection}

\author{Evan A.~Cryer-Jenkins\,\orcidlink{0000-0003-2549-0280}}
\thanks{These authors contributed equally to this work and are listed alphabetically.}
\affiliation{Quantum Measurement Lab, Blackett Laboratory, Imperial College London, London SW7 2BW, United Kingdom}
\author{Kyle D.~Major\,\orcidlink{0000-0002-3268-6946}}
\thanks{These authors contributed equally to this work and are listed alphabetically.}
\affiliation{Quantum Measurement Lab, Blackett Laboratory, Imperial College London, London SW7 2BW, United Kingdom}
\author{Jack Clarke\,\orcidlink{0000-0001-8055-449X}}
\affiliation{Quantum Measurement Lab, Blackett Laboratory, Imperial College London, London SW7 2BW, United Kingdom}
\author{Georg Enzian\,\orcidlink{0000-0002-2603-2874}}
\affiliation{Quantum Measurement Lab, Blackett Laboratory, Imperial College London, London SW7 2BW, United Kingdom}
\affiliation{Clarendon Laboratory, Department of Physics, University of Oxford, OX1 3PU, United Kingdom}
\author{Magdalena Szczykulska\,\orcidlink{0000-0002-5820-7093}}
\affiliation{Clarendon Laboratory, Department of Physics, University of Oxford, OX1 3PU, United Kingdom}
\author{Jinglei Zhang}
\affiliation{Institute for Quantum Computing, University of Waterloo, Waterloo, Ontario, N2L 3G1, Canada}
\affiliation{Department of Physics \& Astronomy, University of Waterloo, Waterloo, Ontario, N2L 3G1, Canada}
\author{Arjun Gupta\,\orcidlink{0009-0009-2994-9125}}
\affiliation{Quantum Measurement Lab, Blackett Laboratory, Imperial College London, London SW7 2BW, United Kingdom}
\author{Anthony C.~Leung\,\orcidlink{0000-0001-9420-292X}}
\affiliation{Quantum Measurement Lab, Blackett Laboratory, Imperial College London, London SW7 2BW, United Kingdom}
\author{Harsh Rathee\,\orcidlink{0009-0006-7762-8543}}
\affiliation{Quantum Measurement Lab, Blackett Laboratory, Imperial College London, London SW7 2BW, United Kingdom}
\author{Andreas {\O}.~Svela\,\orcidlink{0000-0002-3534-3324}}
\affiliation{Quantum Measurement Lab, Blackett Laboratory, Imperial College London, London SW7 2BW, United Kingdom}
\author{Anthony K.~C.~Tan\,\orcidlink{0000-0002-1324-4376}}
\affiliation{Quantum Measurement Lab, Blackett Laboratory, Imperial College London, London SW7 2BW, United Kingdom}
\author{Almut Beige\,\orcidlink{0000-0001-7230-4220}}
\affiliation{The School of Physics and Astronomy, University of Leeds, Leeds LS2 9JT, United Kingdom}
\author{Klaus M{\o}lmer\,\orcidlink{0000-0002-2372-869X}}
\affiliation{Niels Bohr Institute, University of Copenhagen, Blegdamsvej 17, 2100 Copenhagen, Denmark}
\author{Michael R.~Vanner\,\orcidlink{0000-0001-9816-5994}}
\email{m.vanner@imperial.ac.uk}
\homepage{www.qmeas.net}
\affiliation{Quantum Measurement Lab, Blackett Laboratory, Imperial College London, London SW7 2BW, United Kingdom}

\date{\today}

\begin{abstract}
Throughout quantum science and technology, measurement is used as a powerful resource for nonlinear operations and quantum state engineering. In particular, single-photon detection is commonly employed for quantum-information applications and tests of fundamental physics. By contrast, and perhaps counter-intuitively, measurement of the absence of photons also provides useful information, and offers significant potential for a wide range of new experimental directions. Here, we propose and experimentally demonstrate cooling of a mechanical resonator below its laser-cooled mechanical occupation via zero-photon detection on the anti-Stokes scattered optical field and verify this cooling through heterodyne measurements. Our measurements are well captured by a stochastic master equation and the techniques introduced here open new avenues for cooling, quantum thermodynamics, quantum state engineering, and quantum measurement and control.
\end{abstract}


\maketitle

\textit{Introduction.}---Cavity optomechanical laser cooling provides a rich avenue for research and enables the preparation of low-entropy initial states of mechanical oscillators. Building on the pioneering work by Braginsky and colleagues~\cite{braginski1967ponderomotive}, the thermal ground state has now been achieved via laser cooling in the optical~\cite{Chan2011} and microwave~\cite{Teufel2011} domains. These achievements fuelled numerous subsequent developments and laser cooling remains a very active area in cavity optomechanics~\cite{aspelmeyer2014cavity}. Key to the performance of laser cooling is operation in the resolved sideband regime, where the cavity decay rate is much smaller than the mechanical angular frequency, i.e. $\kappa < \omega_m$, to resonantly select the light-mechanics beam-splitter interaction. Outside the resolved-sideband regime, i.e. $\kappa > \omega_m$, optical measurement-based techniques for cooling have been explored with a prominent example being feedback cooling~\cite{mancini1998optomechanical,cohadon1999cooling}, which has also now experimentally reached the thermal ground state~\cite{rossi2018measurement}. An undesirable consequence of both laser cooling and feedback cooling is the mechanical damping rate increases with increasing cooling. Other techniques for cooling utilize pulsed measurement approaches for mechanical position measurements beyond the standard quantum limit~\cite{braginskii1978optimal}, which also enable mechanical squeezing and tomography~\cite{vanner2011pulsed}. Pulsed experiments employing these techniques are also progressing towards measurement-based cooling to the mechanical quantum noise level~\cite{vanner2013cooling,muhonen2019state}.

In quantum optics more broadly, measurement plays a central role for quantum-state engineering and quantum-information applications. Prominent examples include heralded single-photon generation via single-photon detection~\cite{Hong1986}, single-photon addition and subtraction operations~\cite{Ourjoumtsev2006, neergaardNielsen2006, Zavatta2007}, and noiseless linear amplification~\cite{Xiang2010}. While the majority of photon-counting schemes for state engineering utilise the presence of one or more photons, detecting the absence of photons may also modify a state. Such zero-photon detection has been considered in quantum optics as a tool for noiseless attenuation~\cite{Mivcuda2012,Gagatsos2014,Brewster2017,Ye2019}, for Gaussification of entanglement distillation outputs~\cite{Browne2003}, for covert information sharing~\cite{zanforlin2023covert}, and for optical state engineering and reconstruction~\cite{banaszek1996direct,kim1997quasiprobability,Zambra2005}. The statistics of zero-photon events, often discounted in numerical and experimental protocols, also provide useful information for parameter estimation protocols~\cite{Clark2019}, and quantum simulation~\cite{Wein2024}. Recent experimental works have also demonstrated that zero-photon detection after a beam-splitter interaction modifies the output optical field depending on the input photon statistics~\cite{Nunn2021, Nunn2022}.

Quantum measurement techniques utilizing photon counting have also been recently employed in optomechanics applications~\cite{borkje2011proposal,Vanner2013} where correlations between Stokes-scattered and anti-Stokes-scattered fields have been explored~\cite{Lee2012,riedinger2016non}, single- and multi-phonon addition and subtraction operations to thermal mechanical states have been performed~\cite{Enzian2021,Enzian2021_2,Patel2021}, and higher-order phonon correlations have been measured~\cite{Patil2022}. With the growing interest in optomechanical quantum measurement and with the continued interest in laser cooling in a wide range of optomechanical and Brillouin-scattering-based systems~\cite{Qiu2020,Peterson2016,Blazquez2024}, a measurement-based scheme capable of cooling mechanical oscillators further than the limits of laser cooling therefore constitutes a valuable new tool.

In this Letter, we propose and experimentally demonstrate enhanced cooling of a mechanical oscillator below its laser-cooled occupation via zero-photon detection of the anti-Stokes-scattered optical field. The enhancement to the cooling is increased with longer measurement times and is experimentally demonstrated for both a single time-step and a sequence of time-steps. The dynamics of the coupled optomechanical system, probed via heterodyne measurements of the anti-Stokes light, are well captured by a stochastic master equation, which allows the state of the acoustic mode to be determined from the measured light. The zero-photon-detection-enhanced laser cooling is also experimentally contrasted with the case of single-photon detection, which increases the mean mechanical occupation~\cite{Enzian2021}. The effect of detection efficiency on both forms of measurement is explored, showing the reduction in achievable cooling via zero-photon detection as optical loss increases. The enhanced cooling techniques demonstrated here can be applied to many physical systems including optomechanics, electromechanics, atomic spin ensembles, and superconducting circuits, and offers new tools to explore the interface of quantum mechanics and thermodynamics, and perform  quantum measurement and control.

\begin{figure}[t]
\includegraphics[width=\linewidth]{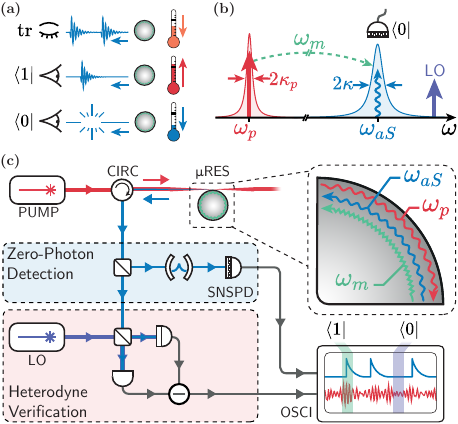}
\caption{\label{fig:setup}Scheme and experimental schematic for zero-photon-measurement-enhanced laser cooling. (a) Cartoon of the effect of ``no measurement'' (corresponding to a trace operation), single-photon detection, and zero-photon detection on a thermal state. (b) The optical mode structure experimentally used to drive the Brillouin interaction. The pump mode is shown in red and the anti-Stokes mode in blue (offset by the Brillouin frequency in green). A local oscillator (LO) is located close in frequency to the anti-Stokes light. (c) Experimental schematic. A pump laser is coupled into a whispering-gallery-mode microresonator (\textmu RES) and the backscattered anti-Stokes light is separated via an optical circulator (CIRC). The signal is then split in two: one portion is directed onto a superconducting nanowire single-photon detector (SNSPD) to perform zero-photon detection and heterodyne measurements are performed in a verification arm. The SNSPD and heterodyne signals are recorded on an oscilloscope (OSCI). Inset: A representation of the pump and counter-propagating anti-Stokes and mechanical (green) modes of the resonator.
}
\end{figure}

\textit{Enhanced Mechanical Cooling via Zero-Photon Detection.}---To laser cool a mechanical oscillator, the optomechanical beamsplitter interaction is brought into resonance and the output frequency-upshifted light is unmonitored, i.e. a partial trace operation. To understand the cooling and heating effects described below, we note that this trace operation can be viewed as an average over detection of all photon numbers, including {zero-,} single-, and multi-photon detection events. As has been recently experimentally demonstrated, the detection of a single photon following the optomechanical beam-splitter interaction is a single-phonon subtraction operation that doubles the mean occupation of an initial thermal state~\cite{Enzian2021, Enzian2021_2, Patel2021}, and multi-phonon subtraction further increases the mean occupation~\cite{Enzian2021_2}. Thus, as is represented schematically in Fig.~\hyperref[fig:setup]{1(a)}, the zero-photon-detection events must constitute a cooling so laser cooling is achieved when all outcomes in the trace are averaged. Therefore, when the output light mode is instead monitored and the zero-photon-detection events are recorded, the cooling is enhanced beyond standard laser cooling.
To see this quantitatively, we model the optomechanical interaction using the beamsplitter Hamiltonian $H_{aS}=\hbar G (a b^{\dagger} + a^{\dagger} b)$, where $a$ and $b$ are the annihilation operators of the optical anti-Stokes and mechanical modes, respectively, and $G$ is the linearized optomechanical coupling rate.
The detection of $n$ photons after a short anti-Stokes interaction of duration $\tau$ is described by the measurement operator $\Upsilon_{n}^{(aS)}=\bra{n}e^{-i H_{aS}\,\tau/\hbar}\ket{0}\propto\left[\cos(G\tau)\right]^{\bd b}b^{n}$. Looking at the two parts to this measurement operator, $b^n$ describes $n$-phonon subtraction and, crucially to this work, the term $\left[\cos(G\tau)\right]^{\bd b}$ obtained for zero-photon detection, when applied to a thermal state yields a new thermal state with a reduced thermal occupation.

In order to model the interaction, open-system dynamics, and measurements, we employ a stochastic master equation (SME) approach.
By considering a series of short interactions and photon-counting measurements, and taking the limit $\tau\rightarrow d{t}$ allows one to derive the SME
\begin{align}
    d{\rho}=&-\frac{i}{\hbar}[H_{aS},\rho]{dt}+\mathcal{G}[a]\rho{dN}-{\eta}\kappa_{ex}\mathcal{H}\!\left[a^{\dagger}a\right]\rho{dt}\nonumber\\
    &+2\kappa_{ex}(1-{\eta})\mathcal{D}\!\left[a\right]\rho{dt}\nonumber+2\kappa_{in}\mathcal{D}\!\left[a\right]\rho{dt}\nonumber\\
    &+2\gamma(\bar{N}+1)\mathcal{D}[b]\rho{dt}+2\gamma\bar{N}\mathcal{D}[b^{\dagger}]\rho{dt}.\label{eq:SME}
\end{align}
Here, ${\eta}$ is the total optical detection efficiency from outside the cavity to the photon counter, $\kappa_{ex} (\kappa_{in})$ is the external (intrinsic) amplitude coupling (decay) rate of the cavity mode, $\gamma$ is the amplitude decay rate of the mechanical mode, $\bar{N}$ is the occupation of the mechanical thermal environment, and the superoperators are given by $\mathcal{G}[O]\rho=O\rho{O}^{\dagger}/\expval{O^{\dagger}O}-\rho$, $\mathcal{H}[O]\rho=O\rho+\rho{O^{\dag}}-\expval{O+O^{\dagger}}\rho$, and $\mathcal{D}[O]\rho=O\rho{O^{\dag}}-\frac{1}{2}\{O^{\dag}O,\rho\}$. Further, the stochastic increment ${dN}=0$ or $1$ for a zero- or single-photon detection event and so ${dN}^2={dN}$. For more details and further theoretical studies using this SME, see the companion paper to this work, Ref.~\cite{cryer2023mechanical}.

\textit{Experimental Setup.}---We implement and verify the effect of zero-photon detection using Brillouin scattering in a $280\,\mu$m diameter fused-silica microsphere resonator operating at room temperature and coupled using a tapered optical fiber. As illustrated in Fig.~\hyperref[fig:setup]{1(b)}, we utilize two cavity modes spaced by the acoustic frequency $\omega_m/2\pi=10.85$\,GHz to select and resonantly drive the the anti-Stokes process. Utilizing a pair of cavity modes in this manner also ensures that the Stokes process is suppressed.

A schematic of the optical setup is shown in Fig.~\hyperref[fig:setup]{1(c)}. Light from a continuous-wave pump laser at $1550$\,nm is sent through a circulator and is evanescently coupled into the microresonator. The transmitted pump field is used to characterise the optical modes and lock the pump laser to a cavity resonance via Pound-Drever-Hall frequency stabilization. The pump cavity mode used in this work has a linewidth of $2\kappa_{p}/2\pi=6.1$\,MHz and the anti-Stokes mode has a linewidth $2\kappa/2\pi=45.2$\,MHz, with an external coupling rate of $2\kappa_{ex}/2\pi=10.1$\,MHz, full-widths-at-half-maxima. The mechanical linewidth is $2\gamma/2\pi=45$\,MHz, which is in good agreement with previous measurements and bulk decay rates in amorphous silica~\cite{Boyd1992,Enzian2019}. The pump-enhanced coupling rate is approximately $G/2\pi=3.3$\,MHz for an input pump power of 0.37\,mW, yielding a modest cooperativity of $C=G^2/\kappa\gamma = 0.02$. The photon counting detection efficiency from the cavity output, including optical losses, is up to $\eta=0.32\%$ depending on the filtering configuration, which is consistent with SME simulations.

The backscattered anti-Stokes light couples back into the tapered fiber, is separated from the counter-propagating pump by a circulator, and then split so that half the light is measured by a superconducting nanowire single-photon detector (SNSPD) following spectral filtering, and the remaining half is measured via heterodyne detection.
Low optical powers were utilised in this experiment to avoid multiphoton detection by the SNSPD, which recorded an average count rate of $\sim10^5$\,s\textsuperscript{-1}.
The heterodyne and SNSPD signals are recorded on an oscilloscope with a 1.25\,GHz sampling rate. The nanowire rise time is $\sim100$\,ps and the oscilloscope sampling time is 0.8\,ns, which are shorter than all relevant timescales in this experiment so photon counting well approximates an instantaneous measurement.
We record $7.5\times10^{5}$ heterodyne and corresponding SNSPD time traces of 200~ns duration. We then determine the heterodyne statistics conditioned on the SNSPD measurement record. We'd like to clarify here that the SME is not used for data analysis but rather gives a theoretical prediction for what our experiment observes. Measurement over a single time sample and a string of multiple time samples, i.e. an SNSPD measurement record, are examined.

\begin{figure*}
    \centering
    \includegraphics[width=\textwidth]{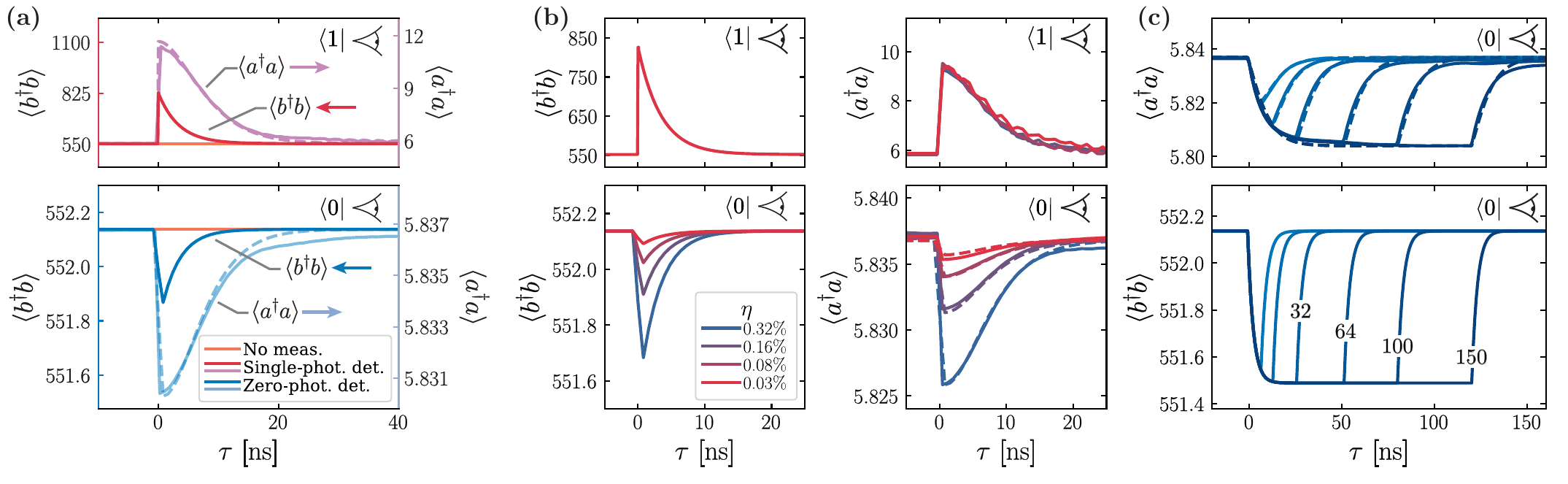}
    \caption{Experimental observations of zero-photon-detection-enhanced laser cooling and dynamics associated with the photon-counting measurement record.
    (a) Plots of the mechanical and optical mean occupations for single-photon detection (upper row) and zero-photon detection (lower row) over a single time step as a function of time $\tau$ from the measurement. The left vertical axes are the mechanical mean occupations and the right vertical axes are the optical mean occupations.
    (b) Plots illustrating the impact of efficiency on the single-photon and zero-photon cases for a range of efficiencies implemented by including additional attenuation. The scaling with efficiency for zero-photon detection is observed whereas the single-photon case remains unaffected and the four curves overlap.
    (c) Enhanced mechanical cooling through continuous zero-photon detection. Plot of the intracavity mean photon number (upper row) obtained from the experimental heterodyne data (solid lines) overlayed with the prediction made by the SME (dashed lines) using the experimental parameters summarized in~\cite{supp}. The corresponding mechanical occupation (lower row) showing the improved cooling performance with measurement time. The lengths of the zero-photon detection strings used are 8, 16, 32, 64, 100 and 150 (in units of 0.8\,ns). An efficiency of 0.19\% is used in plots (a) and (c), and the efficiencies for (b) are given in the legend.}
    \label{fig:fig2}
\end{figure*}
 
\textit{Results and Discussion.}---Fig.~\hyperref[fig:fig2]{2(a)} shows the mean phonon number $\expval{b^{\dagger}b}$ and intracavity photon number $\expval{a^{\dagger}a}$ obtained from the heterodyne signals for the three cases of ``no measurement", single-photon detection, and zero-photon detection over a single time sample. The no-measurement trace shows occupations that are constant in time, equal to the laser-cooling steady-state occupations. The upper traces corresponding to single-photon detection demonstrate an approximate doubling in the optical occupation and an accompanying increase in the mean number of phonons. As this experiment does not operate in the adiabatic regime where $\kappa_{ex}$ is larger than all other relevant rates, the increase in mean phonon number is less than a factor of 2. The small deviation from optical doubling is attributed to imperfect filtering of backscattered pump light and mode mismatch between the two detection arms. The lower plots depicting zero-photon detection illustrate a reduction from the laser-cooling occupations, demonstrating the enhancement beyond laser cooling.
The experimental dynamics observed are in good agreement with the prediction made by the SME, which is also equivalent to the measurement-operator approach for this case.

The impact of detection efficiency on zero-photon-measurement-enhanced laser cooling is examined by including optical attenuators prior to the SNSPD. For a single time sample, the depth of the enhanced cooling observed (c.f. Fig.~\hyperref[fig:fig2]{2(b)}) scales linearly with $\eta$ as is predicted by our model~\cite{supp}.
By contrast, in the limit of low dark-counts, the change in the state following single-photon detection does not depend on the detection efficiency, rather, this just affects the heralding probability.
Note, a higher efficiency is employed for the measurement in Fig.~\hyperref[fig:fig2]{2(b)} than in Fig.~\hyperref[fig:fig2]{2(a,~c)} by using less spectral filtering. This approach improves the performance of the zero-photon-detection-enhanced laser cooling but degrades the quality of the single-photon measurement as there are more false positives due to deleterious back-scattered pump light.

Beyond zero-photon detection over a single time sample, this measurement-based cooling strategy may be further improved by implementing continuous zero-photon detection as illustrated in Fig.~\hyperref[fig:fig2]{2(c)}. 
Strings of up to 150 consecutive zero-photon detection events are considered (each lasting 0.8~ns), illustrating an improved cooling up to a limit where heating rates balance information extracted from the system. Performing this continuous-time zero-photon detection for the modest cooperativity in this experiment, the mechanical occupation was reduced by a further 11.1\,\% beyond laser cooling, which is heralded with a 98\% probability for the 32 string case and 92\% probability for the 150 string case.
For a given efficiency, the limit of cooling via continuous zero-photon detection is reached for a strongly overcoupled cavity, i.e. $\kappa=\kappa_{ex}$. When these conditions are met, the measurement-enhanced laser cooling gives a mechanical occupation of $\bar{n}_{lim} =\left[{-(1+C)+\sqrt{(1+C)^2+4\eta\bar{N} C}}\right]/{2\eta C}$~\cite{supp}
. In this regime, taking $C=0.02$ and $\bar{N}=558$ from our experiment, the laser-cooled occupation is $\bar{n}_{LC}=\bar{N}/(1+C)=547$ and zero-photon-detection-enhanced occupation reduces to $\bar{n}_{lim}=143$ as $\eta\rightarrow1$, which corresponds to $26\%$ of the laser-cooled occupation.

\textit{Conclusions and Outlook.}---We propose and experimentally demonstrate the enhancement of laser cooling of a mechanical mode via zero-photon detection of the frequency-upshifted optical field. The enhanced cooling is well described by a stochastic master equation, which captures the enhancement with a sequence of time-steps of the measurement. The effects of detection inefficiency are understood and experimentally explored, as well as being contrasted against the loss-resilient effects of single-photon detection.
The final occupation of the enhanced laser cooled state scales favourably with cooperativity and detection efficiency, and thus, improvements to these quantities result in a greatly improved cooling performance. 
For instance, for lower mechanical-frequency systems with larger thermal populations such as mechanical membranes~\cite{Kristensen2024} and forward Brillouin scattering~\cite{Bahl2012}, the effect is also significantly increased and can represent a significant advantage over conventional laser cooling alone for practical heralding probabilities.
More generally, for high efficiency and cooperativity, to reduce the mean thermal occupation using zero-photon detection below a target value $\bar{n}_{*}$ requires that the laser-cooled state has an occupation below $\bar{n}_{LC}=\bar{n}_{*}^2+\bar{n}_{*}$. Indeed, to achieve $\bar{n}_{*}<1$ one requires laser cooling to $\bar{n}_{LC}<2$, which is now achieved by several experimental platforms including optomechanical crystals~\cite{Chan2011, Qiu2020}, toroidal microresonators~\cite{Verhagen2012}, and membranes~\cite{Peterson2016}. Taking the parameters from Ref.~\cite{Verhagen2012} as an example, which performed laser cooling from 197 to 1.7 mean thermal phonons, employing zero-photon-detection in that experiment could have enabled a mean thermal occupation of 0.90 to be reached. 
Such reductions to the laser-cooled mechanical occupations at these low levels, even for $\bar{n}_{LC}<1$, are especially valuable for quantum state engineering protocols that are sensitive to initial thermal occupation.

This work utilizes the counter-intuitive fact that a measurement of ``nothing" can have a significant impact to the state of a physical system. The measurement-based technique introduced here can be readily employed by several experimental systems to enhance their laser cooling performance, and more immediately performed in systems and protocols where photon-detection is already employed~\cite{borkje2011proposal,Vanner2013,Lee2012, cohen2015phonon, riedinger2016non, Enzian2021, Enzian2021_2, Patel2021, Patil2022, cryer2023second, Milburn2016, galinskiy2023non}. This technique will also be of particular value to experiments aiming to achieve ultra-low thermal occupations from room temperature, which is a current goal within optomechanics.
We'd also like to highlight that the additional cooling provided by this technique does not degrade the mechanical linewidth, further increasing the versatility of this approach.
Beyond enhancing laser cooling, the zero-photon-detection-based technique introduced here expands the toolset of measurement-based operations to mechanical states, provides a method to perform mechanical noiseless attenuation, and opens new avenues for studies of quantum thermodynamics, and quantum measurement and control.

\begin{acknowledgments}
\textit{Acknowledgements.}---We acknowledge useful discussions with Lewis A. Clark, Rufus Clarke, Artie Clayton-Major,  Lars Freisem, Daniel Hodgson, Gerard J. Milburn, and John J. Price. This project was supported by UK Research and Innovation (MR/S032924/1, MR/X024105/1), the Engineering and Physical Sciences Research Council (EP/T031271/1, EP/P510257/1), the Science and Technology Facilities Council (ST/W006553/1), the Royal Society, and the Aker Scholarship.
\end{acknowledgments}


\begin{thebibliography}{51}%
\makeatletter
\providecommand \@ifxundefined [1]{%
 \@ifx{#1\undefined}
}%
\providecommand \@ifnum [1]{%
 \ifnum #1\expandafter \@firstoftwo
 \else \expandafter \@secondoftwo
 \fi
}%
\providecommand \@ifx [1]{%
 \ifx #1\expandafter \@firstoftwo
 \else \expandafter \@secondoftwo
 \fi
}%
\providecommand \natexlab [1]{#1}%
\providecommand \enquote  [1]{``#1''}%
\providecommand \bibnamefont  [1]{#1}%
\providecommand \bibfnamefont [1]{#1}%
\providecommand \citenamefont [1]{#1}%
\providecommand \href@noop [0]{\@secondoftwo}%
\providecommand \href [0]{\begingroup \@sanitize@url \@href}%
\providecommand \@href[1]{\@@startlink{#1}\@@href}%
\providecommand \@@href[1]{\endgroup#1\@@endlink}%
\providecommand \@sanitize@url [0]{\catcode `\\12\catcode `\$12\catcode `\&12\catcode `\#12\catcode `\^12\catcode `\_12\catcode `\%12\relax}%
\providecommand \@@startlink[1]{}%
\providecommand \@@endlink[0]{}%
\providecommand \url  [0]{\begingroup\@sanitize@url \@url }%
\providecommand \@url [1]{\endgroup\@href {#1}{\urlprefix }}%
\providecommand \urlprefix  [0]{URL }%
\providecommand \Eprint [0]{\href }%
\providecommand \doibase [0]{https://doi.org/}%
\providecommand \selectlanguage [0]{\@gobble}%
\providecommand \bibinfo  [0]{\@secondoftwo}%
\providecommand \bibfield  [0]{\@secondoftwo}%
\providecommand \translation [1]{[#1]}%
\providecommand \BibitemOpen [0]{}%
\providecommand \bibitemStop [0]{}%
\providecommand \bibitemNoStop [0]{.\EOS\space}%
\providecommand \EOS [0]{\spacefactor3000\relax}%
\providecommand \BibitemShut  [1]{\csname bibitem#1\endcsname}%
\let\auto@bib@innerbib\@empty
\bibitem [{\citenamefont {Braginski}\ and\ \citenamefont {Manukin}(1967)}]{braginski1967ponderomotive}%
  \BibitemOpen
  \bibfield  {author} {\bibinfo {author} {\bibfnamefont {V.}~\bibnamefont {Braginski}}\ and\ \bibinfo {author} {\bibfnamefont {A.}~\bibnamefont {Manukin}},\ }\href@noop {} {\bibfield  {journal} {\bibinfo  {journal} {Soviet Physics JETP}\ }\textbf {\bibinfo {volume} {25}},\ \bibinfo {pages} {653} (\bibinfo {year} {1967})}\BibitemShut {NoStop}%
\bibitem [{\citenamefont {Chan}\ \emph {et~al.}(2011)\citenamefont {Chan}, \citenamefont {Alegre}, \citenamefont {Safavi-Naeini}, \citenamefont {Hill}, \citenamefont {Krause}, \citenamefont {Gr{\"o}blacher}, \citenamefont {Aspelmeyer},\ and\ \citenamefont {Painter}}]{Chan2011}%
  \BibitemOpen
  \bibfield  {author} {\bibinfo {author} {\bibfnamefont {J.}~\bibnamefont {Chan}}, \bibinfo {author} {\bibfnamefont {T.~M.}\ \bibnamefont {Alegre}}, \bibinfo {author} {\bibfnamefont {A.~H.}\ \bibnamefont {Safavi-Naeini}}, \bibinfo {author} {\bibfnamefont {J.~T.}\ \bibnamefont {Hill}}, \bibinfo {author} {\bibfnamefont {A.}~\bibnamefont {Krause}}, \bibinfo {author} {\bibfnamefont {S.}~\bibnamefont {Gr{\"o}blacher}}, \bibinfo {author} {\bibfnamefont {M.}~\bibnamefont {Aspelmeyer}},\ and\ \bibinfo {author} {\bibfnamefont {O.}~\bibnamefont {Painter}},\ }\href {https://www.nature.com/articles/nature10461} {\bibfield  {journal} {\bibinfo  {journal} {Nature}\ }\textbf {\bibinfo {volume} {478}},\ \bibinfo {pages} {89} (\bibinfo {year} {2011})}\BibitemShut {NoStop}%
\bibitem [{\citenamefont {Teufel}\ \emph {et~al.}(2011)\citenamefont {Teufel}, \citenamefont {Donner}, \citenamefont {Li}, \citenamefont {Harlow}, \citenamefont {Allman}, \citenamefont {Cicak}, \citenamefont {Sirois}, \citenamefont {Whittaker}, \citenamefont {Lehnert},\ and\ \citenamefont {Simmonds}}]{Teufel2011}%
  \BibitemOpen
  \bibfield  {author} {\bibinfo {author} {\bibfnamefont {J.~D.}\ \bibnamefont {Teufel}}, \bibinfo {author} {\bibfnamefont {T.}~\bibnamefont {Donner}}, \bibinfo {author} {\bibfnamefont {D.}~\bibnamefont {Li}}, \bibinfo {author} {\bibfnamefont {J.~W.}\ \bibnamefont {Harlow}}, \bibinfo {author} {\bibfnamefont {M.}~\bibnamefont {Allman}}, \bibinfo {author} {\bibfnamefont {K.}~\bibnamefont {Cicak}}, \bibinfo {author} {\bibfnamefont {A.~J.}\ \bibnamefont {Sirois}}, \bibinfo {author} {\bibfnamefont {J.~D.}\ \bibnamefont {Whittaker}}, \bibinfo {author} {\bibfnamefont {K.~W.}\ \bibnamefont {Lehnert}},\ and\ \bibinfo {author} {\bibfnamefont {R.~W.}\ \bibnamefont {Simmonds}},\ }\href {https://www.nature.com/articles/nature10261} {\bibfield  {journal} {\bibinfo  {journal} {Nature}\ }\textbf {\bibinfo {volume} {475}},\ \bibinfo {pages} {359} (\bibinfo {year} {2011})}\BibitemShut {NoStop}%
\bibitem [{\citenamefont {Aspelmeyer}\ \emph {et~al.}(2014)\citenamefont {Aspelmeyer}, \citenamefont {Kippenberg},\ and\ \citenamefont {Marquardt}}]{aspelmeyer2014cavity}%
  \BibitemOpen
  \bibfield  {author} {\bibinfo {author} {\bibfnamefont {M.}~\bibnamefont {Aspelmeyer}}, \bibinfo {author} {\bibfnamefont {T.~J.}\ \bibnamefont {Kippenberg}},\ and\ \bibinfo {author} {\bibfnamefont {F.}~\bibnamefont {Marquardt}},\ }\href {https://doi.org/10.1103/RevModPhys.86.1391} {\bibfield  {journal} {\bibinfo  {journal} {Reviews of Modern Physics}\ }\textbf {\bibinfo {volume} {86}},\ \bibinfo {pages} {1391} (\bibinfo {year} {2014})}\BibitemShut {NoStop}%
\bibitem [{\citenamefont {Mancini}\ \emph {et~al.}(1998)\citenamefont {Mancini}, \citenamefont {Vitali},\ and\ \citenamefont {Tombesi}}]{mancini1998optomechanical}%
  \BibitemOpen
  \bibfield  {author} {\bibinfo {author} {\bibfnamefont {S.}~\bibnamefont {Mancini}}, \bibinfo {author} {\bibfnamefont {D.}~\bibnamefont {Vitali}},\ and\ \bibinfo {author} {\bibfnamefont {P.}~\bibnamefont {Tombesi}},\ }\href {https://doi.org/10.1103/PhysRevLett.80.688} {\bibfield  {journal} {\bibinfo  {journal} {Physical Review Letters}\ }\textbf {\bibinfo {volume} {80}},\ \bibinfo {pages} {688} (\bibinfo {year} {1998})}\BibitemShut {NoStop}%
\bibitem [{\citenamefont {Cohadon}\ \emph {et~al.}(1999)\citenamefont {Cohadon}, \citenamefont {Heidmann},\ and\ \citenamefont {Pinard}}]{cohadon1999cooling}%
  \BibitemOpen
  \bibfield  {author} {\bibinfo {author} {\bibfnamefont {P.~F.}\ \bibnamefont {Cohadon}}, \bibinfo {author} {\bibfnamefont {A.}~\bibnamefont {Heidmann}},\ and\ \bibinfo {author} {\bibfnamefont {M.}~\bibnamefont {Pinard}},\ }\href {https://doi.org/10.1103/PhysRevLett.83.3174} {\bibfield  {journal} {\bibinfo  {journal} {Physical Review Letters}\ }\textbf {\bibinfo {volume} {83}},\ \bibinfo {pages} {3174} (\bibinfo {year} {1999})}\BibitemShut {NoStop}%
\bibitem [{\citenamefont {Rossi}\ \emph {et~al.}(2018)\citenamefont {Rossi}, \citenamefont {Mason}, \citenamefont {Chen}, \citenamefont {Tsaturyan},\ and\ \citenamefont {Schliesser}}]{rossi2018measurement}%
  \BibitemOpen
  \bibfield  {author} {\bibinfo {author} {\bibfnamefont {M.}~\bibnamefont {Rossi}}, \bibinfo {author} {\bibfnamefont {D.}~\bibnamefont {Mason}}, \bibinfo {author} {\bibfnamefont {J.}~\bibnamefont {Chen}}, \bibinfo {author} {\bibfnamefont {Y.}~\bibnamefont {Tsaturyan}},\ and\ \bibinfo {author} {\bibfnamefont {A.}~\bibnamefont {Schliesser}},\ }\href {https://doi.org/10.1038/s41586-018-0643-8} {\bibfield  {journal} {\bibinfo  {journal} {Nature}\ }\textbf {\bibinfo {volume} {563}},\ \bibinfo {pages} {53} (\bibinfo {year} {2018})}\BibitemShut {NoStop}%
\bibitem [{\citenamefont {Braginsky}\ \emph {et~al.}(1978)\citenamefont {Braginsky}, \citenamefont {Vorontsov},\ and\ \citenamefont {Khalili}}]{braginskii1978optimal}%
  \BibitemOpen
  \bibfield  {author} {\bibinfo {author} {\bibfnamefont {V.~B.}\ \bibnamefont {Braginsky}}, \bibinfo {author} {\bibfnamefont {Y.~I.}\ \bibnamefont {Vorontsov}},\ and\ \bibinfo {author} {\bibfnamefont {F.~Y.}\ \bibnamefont {Khalili}},\ }\href@noop {} {\bibfield  {journal} {\bibinfo  {journal} {JETP Letters}\ }\textbf {\bibinfo {volume} {27}} (\bibinfo {year} {1978})}\BibitemShut {NoStop}%
\bibitem [{\citenamefont {Vanner}\ \emph {et~al.}(2011)\citenamefont {Vanner}, \citenamefont {Pikovski}, \citenamefont {Cole}, \citenamefont {Kim}, \citenamefont {Brukner}, \citenamefont {Hammerer}, \citenamefont {Milburn},\ and\ \citenamefont {Aspelmeyer}}]{vanner2011pulsed}%
  \BibitemOpen
  \bibfield  {author} {\bibinfo {author} {\bibfnamefont {M.~R.}\ \bibnamefont {Vanner}}, \bibinfo {author} {\bibfnamefont {I.}~\bibnamefont {Pikovski}}, \bibinfo {author} {\bibfnamefont {G.~D.}\ \bibnamefont {Cole}}, \bibinfo {author} {\bibfnamefont {M.~S.}\ \bibnamefont {Kim}}, \bibinfo {author} {\bibfnamefont {{\v{C}}.}~\bibnamefont {Brukner}}, \bibinfo {author} {\bibfnamefont {K.}~\bibnamefont {Hammerer}}, \bibinfo {author} {\bibfnamefont {G.~J.}\ \bibnamefont {Milburn}},\ and\ \bibinfo {author} {\bibfnamefont {M.}~\bibnamefont {Aspelmeyer}},\ }\href {https://doi.org/10.1073/pnas.1105098108} {\bibfield  {journal} {\bibinfo  {journal} {Proceedings of the National Academy of Sciences}\ }\textbf {\bibinfo {volume} {108}},\ \bibinfo {pages} {16182} (\bibinfo {year} {2011})}\BibitemShut {NoStop}%
\bibitem [{\citenamefont {Vanner}\ \emph {et~al.}(2013{\natexlab{a}})\citenamefont {Vanner}, \citenamefont {Hofer}, \citenamefont {Cole},\ and\ \citenamefont {Aspelmeyer}}]{vanner2013cooling}%
  \BibitemOpen
  \bibfield  {author} {\bibinfo {author} {\bibfnamefont {M.~R.}\ \bibnamefont {Vanner}}, \bibinfo {author} {\bibfnamefont {J.}~\bibnamefont {Hofer}}, \bibinfo {author} {\bibfnamefont {G.~D.}\ \bibnamefont {Cole}},\ and\ \bibinfo {author} {\bibfnamefont {M.}~\bibnamefont {Aspelmeyer}},\ }\href {https://doi.org/10.1038/ncomms3295} {\bibfield  {journal} {\bibinfo  {journal} {Nature Communications}\ }\textbf {\bibinfo {volume} {4}},\ \bibinfo {pages} {2295} (\bibinfo {year} {2013}{\natexlab{a}})}\BibitemShut {NoStop}%
\bibitem [{\citenamefont {Muhonen}\ \emph {et~al.}(2019)\citenamefont {Muhonen}, \citenamefont {La~Gala}, \citenamefont {Leijssen},\ and\ \citenamefont {Verhagen}}]{muhonen2019state}%
  \BibitemOpen
  \bibfield  {author} {\bibinfo {author} {\bibfnamefont {J.~T.}\ \bibnamefont {Muhonen}}, \bibinfo {author} {\bibfnamefont {G.~R.}\ \bibnamefont {La~Gala}}, \bibinfo {author} {\bibfnamefont {R.}~\bibnamefont {Leijssen}},\ and\ \bibinfo {author} {\bibfnamefont {E.}~\bibnamefont {Verhagen}},\ }\href {https://doi.org/10.1103/PhysRevLett.123.113601} {\bibfield  {journal} {\bibinfo  {journal} {Physical Review Letters}\ }\textbf {\bibinfo {volume} {123}},\ \bibinfo {pages} {113601} (\bibinfo {year} {2019})}\BibitemShut {NoStop}%
\bibitem [{\citenamefont {Hong}\ and\ \citenamefont {Mandel}(1986)}]{Hong1986}%
  \BibitemOpen
  \bibfield  {author} {\bibinfo {author} {\bibfnamefont {C.~K.}\ \bibnamefont {Hong}}\ and\ \bibinfo {author} {\bibfnamefont {L.}~\bibnamefont {Mandel}},\ }\href {https://doi.org/10.1103/PhysRevLett.56.58} {\bibfield  {journal} {\bibinfo  {journal} {Physical Review Letters}\ }\textbf {\bibinfo {volume} {56}},\ \bibinfo {pages} {58} (\bibinfo {year} {1986})}\BibitemShut {NoStop}%
\bibitem [{\citenamefont {Ourjoumtsev}\ \emph {et~al.}(2006)\citenamefont {Ourjoumtsev}, \citenamefont {Tualle-Brouri}, \citenamefont {Laurat},\ and\ \citenamefont {Grangier}}]{Ourjoumtsev2006}%
  \BibitemOpen
  \bibfield  {author} {\bibinfo {author} {\bibfnamefont {A.}~\bibnamefont {Ourjoumtsev}}, \bibinfo {author} {\bibfnamefont {R.}~\bibnamefont {Tualle-Brouri}}, \bibinfo {author} {\bibfnamefont {J.}~\bibnamefont {Laurat}},\ and\ \bibinfo {author} {\bibfnamefont {P.}~\bibnamefont {Grangier}},\ }\href {https://doi.org/10.1126/science.1122858} {\bibfield  {journal} {\bibinfo  {journal} {Science}\ }\textbf {\bibinfo {volume} {312}},\ \bibinfo {pages} {83} (\bibinfo {year} {2006})}\BibitemShut {NoStop}%
\bibitem [{\citenamefont {Neergaard-Nielsen}\ \emph {et~al.}(2006)\citenamefont {Neergaard-Nielsen}, \citenamefont {Nielsen}, \citenamefont {Hettich}, \citenamefont {Mølmer},\ and\ \citenamefont {Polzik}}]{neergaardNielsen2006}%
  \BibitemOpen
  \bibfield  {author} {\bibinfo {author} {\bibfnamefont {J.~S.}\ \bibnamefont {Neergaard-Nielsen}}, \bibinfo {author} {\bibfnamefont {B.~M.}\ \bibnamefont {Nielsen}}, \bibinfo {author} {\bibfnamefont {C.}~\bibnamefont {Hettich}}, \bibinfo {author} {\bibfnamefont {K.}~\bibnamefont {Mølmer}},\ and\ \bibinfo {author} {\bibfnamefont {E.~S.}\ \bibnamefont {Polzik}},\ }\href {https://doi.org/10.1103/PhysRevLett.97.083604} {\bibfield  {journal} {\bibinfo  {journal} {Physical Review Letters}\ }\textbf {\bibinfo {volume} {97}},\ \bibinfo {pages} {083604} (\bibinfo {year} {2006})}\BibitemShut {NoStop}%
\bibitem [{\citenamefont {Zavatta}\ \emph {et~al.}(2007)\citenamefont {Zavatta}, \citenamefont {Parigi},\ and\ \citenamefont {Bellini}}]{Zavatta2007}%
  \BibitemOpen
  \bibfield  {author} {\bibinfo {author} {\bibfnamefont {A.}~\bibnamefont {Zavatta}}, \bibinfo {author} {\bibfnamefont {V.}~\bibnamefont {Parigi}},\ and\ \bibinfo {author} {\bibfnamefont {M.}~\bibnamefont {Bellini}},\ }\href {https://doi.org/10.1103/PhysRevA.75.052106} {\bibfield  {journal} {\bibinfo  {journal} {Physical Review A}\ }\textbf {\bibinfo {volume} {75}},\ \bibinfo {pages} {052106} (\bibinfo {year} {2007})}\BibitemShut {NoStop}%
\bibitem [{\citenamefont {Xiang}\ \emph {et~al.}(2010)\citenamefont {Xiang}, \citenamefont {Ralph}, \citenamefont {Lund}, \citenamefont {Walk},\ and\ \citenamefont {Pryde}}]{Xiang2010}%
  \BibitemOpen
  \bibfield  {author} {\bibinfo {author} {\bibfnamefont {G.-Y.}\ \bibnamefont {Xiang}}, \bibinfo {author} {\bibfnamefont {T.~C.}\ \bibnamefont {Ralph}}, \bibinfo {author} {\bibfnamefont {A.~P.}\ \bibnamefont {Lund}}, \bibinfo {author} {\bibfnamefont {N.}~\bibnamefont {Walk}},\ and\ \bibinfo {author} {\bibfnamefont {G.~J.}\ \bibnamefont {Pryde}},\ }\href {https://www.nature.com/articles/nphoton.2010.35} {\bibfield  {journal} {\bibinfo  {journal} {Nature Photonics}\ }\textbf {\bibinfo {volume} {4}},\ \bibinfo {pages} {316} (\bibinfo {year} {2010})}\BibitemShut {NoStop}%
\bibitem [{\citenamefont {Mi{\v{c}}uda}\ \emph {et~al.}(2012)\citenamefont {Mi{\v{c}}uda}, \citenamefont {Straka}, \citenamefont {Mikov{\'a}}, \citenamefont {Du{\v{s}}ek}, \citenamefont {Cerf}, \citenamefont {Fiur{\'a}{\v{s}}ek},\ and\ \citenamefont {Je{\v{z}}ek}}]{Mivcuda2012}%
  \BibitemOpen
  \bibfield  {author} {\bibinfo {author} {\bibfnamefont {M.}~\bibnamefont {Mi{\v{c}}uda}}, \bibinfo {author} {\bibfnamefont {I.}~\bibnamefont {Straka}}, \bibinfo {author} {\bibfnamefont {M.}~\bibnamefont {Mikov{\'a}}}, \bibinfo {author} {\bibfnamefont {M.}~\bibnamefont {Du{\v{s}}ek}}, \bibinfo {author} {\bibfnamefont {N.~J.}\ \bibnamefont {Cerf}}, \bibinfo {author} {\bibfnamefont {J.}~\bibnamefont {Fiur{\'a}{\v{s}}ek}},\ and\ \bibinfo {author} {\bibfnamefont {M.}~\bibnamefont {Je{\v{z}}ek}},\ }\href {https://doi.org/10.1103/PhysRevLett.109.180503} {\bibfield  {journal} {\bibinfo  {journal} {Physical Review Letters}\ }\textbf {\bibinfo {volume} {109}},\ \bibinfo {pages} {180503} (\bibinfo {year} {2012})}\BibitemShut {NoStop}%
\bibitem [{\citenamefont {Gagatsos}\ \emph {et~al.}(2014)\citenamefont {Gagatsos}, \citenamefont {Fiur{\'a}{\v{s}}ek}, \citenamefont {Zavatta}, \citenamefont {Bellini},\ and\ \citenamefont {Cerf}}]{Gagatsos2014}%
  \BibitemOpen
  \bibfield  {author} {\bibinfo {author} {\bibfnamefont {C.}~\bibnamefont {Gagatsos}}, \bibinfo {author} {\bibfnamefont {J.}~\bibnamefont {Fiur{\'a}{\v{s}}ek}}, \bibinfo {author} {\bibfnamefont {A.}~\bibnamefont {Zavatta}}, \bibinfo {author} {\bibfnamefont {M.}~\bibnamefont {Bellini}},\ and\ \bibinfo {author} {\bibfnamefont {N.}~\bibnamefont {Cerf}},\ }\href {https://doi.org/10.1103/PhysRevA.89.062311} {\bibfield  {journal} {\bibinfo  {journal} {Physical Review A}\ }\textbf {\bibinfo {volume} {89}},\ \bibinfo {pages} {062311} (\bibinfo {year} {2014})}\BibitemShut {NoStop}%
\bibitem [{\citenamefont {Brewster}\ \emph {et~al.}(2017)\citenamefont {Brewster}, \citenamefont {Nodurft}, \citenamefont {Pittman},\ and\ \citenamefont {Franson}}]{Brewster2017}%
  \BibitemOpen
  \bibfield  {author} {\bibinfo {author} {\bibfnamefont {R.}~\bibnamefont {Brewster}}, \bibinfo {author} {\bibfnamefont {I.}~\bibnamefont {Nodurft}}, \bibinfo {author} {\bibfnamefont {T.}~\bibnamefont {Pittman}},\ and\ \bibinfo {author} {\bibfnamefont {J.}~\bibnamefont {Franson}},\ }\href {https://doi.org/10.1103/PhysRevA.96.042307} {\bibfield  {journal} {\bibinfo  {journal} {Physical Review A}\ }\textbf {\bibinfo {volume} {96}},\ \bibinfo {pages} {042307} (\bibinfo {year} {2017})}\BibitemShut {NoStop}%
\bibitem [{\citenamefont {Ye}\ \emph {et~al.}(2019)\citenamefont {Ye}, \citenamefont {Zhong}, \citenamefont {Liao}, \citenamefont {Huang}, \citenamefont {Hu},\ and\ \citenamefont {Guo}}]{Ye2019}%
  \BibitemOpen
  \bibfield  {author} {\bibinfo {author} {\bibfnamefont {W.}~\bibnamefont {Ye}}, \bibinfo {author} {\bibfnamefont {H.}~\bibnamefont {Zhong}}, \bibinfo {author} {\bibfnamefont {Q.}~\bibnamefont {Liao}}, \bibinfo {author} {\bibfnamefont {D.}~\bibnamefont {Huang}}, \bibinfo {author} {\bibfnamefont {L.}~\bibnamefont {Hu}},\ and\ \bibinfo {author} {\bibfnamefont {Y.}~\bibnamefont {Guo}},\ }\href {https://doi.org/10.1364/OE.27.017186} {\bibfield  {journal} {\bibinfo  {journal} {Optics Express}\ }\textbf {\bibinfo {volume} {27}},\ \bibinfo {pages} {17186} (\bibinfo {year} {2019})}\BibitemShut {NoStop}%
\bibitem [{\citenamefont {Browne}\ \emph {et~al.}(2003)\citenamefont {Browne}, \citenamefont {Eisert}, \citenamefont {Scheel},\ and\ \citenamefont {Plenio}}]{Browne2003}%
  \BibitemOpen
  \bibfield  {author} {\bibinfo {author} {\bibfnamefont {D.~E.}\ \bibnamefont {Browne}}, \bibinfo {author} {\bibfnamefont {J.}~\bibnamefont {Eisert}}, \bibinfo {author} {\bibfnamefont {S.}~\bibnamefont {Scheel}},\ and\ \bibinfo {author} {\bibfnamefont {M.~B.}\ \bibnamefont {Plenio}},\ }\href {https://doi.org/10.1103/PhysRevA.67.062320} {\bibfield  {journal} {\bibinfo  {journal} {Physical Review A}\ }\textbf {\bibinfo {volume} {67}},\ \bibinfo {pages} {062320} (\bibinfo {year} {2003})}\BibitemShut {NoStop}%
\bibitem [{\citenamefont {Zanforlin}\ \emph {et~al.}(2023)\citenamefont {Zanforlin}, \citenamefont {Tatsi}, \citenamefont {Jeffers},\ and\ \citenamefont {Buller}}]{zanforlin2023covert}%
  \BibitemOpen
  \bibfield  {author} {\bibinfo {author} {\bibfnamefont {U.}~\bibnamefont {Zanforlin}}, \bibinfo {author} {\bibfnamefont {G.}~\bibnamefont {Tatsi}}, \bibinfo {author} {\bibfnamefont {J.}~\bibnamefont {Jeffers}},\ and\ \bibinfo {author} {\bibfnamefont {G.~S.}\ \bibnamefont {Buller}},\ }\href {https://doi.org/10.1103/PhysRevA.107.022619} {\bibfield  {journal} {\bibinfo  {journal} {Physical Review A}\ }\textbf {\bibinfo {volume} {107}},\ \bibinfo {pages} {022619} (\bibinfo {year} {2023})}\BibitemShut {NoStop}%
\bibitem [{\citenamefont {Banaszek}\ and\ \citenamefont {W{\'o}dkiewicz}(1996)}]{banaszek1996direct}%
  \BibitemOpen
  \bibfield  {author} {\bibinfo {author} {\bibfnamefont {K.}~\bibnamefont {Banaszek}}\ and\ \bibinfo {author} {\bibfnamefont {K.}~\bibnamefont {W{\'o}dkiewicz}},\ }\href {https://doi.org/10.1103/PhysRevLett.76.4344} {\bibfield  {journal} {\bibinfo  {journal} {Physical Review Letters}\ }\textbf {\bibinfo {volume} {76}},\ \bibinfo {pages} {4344} (\bibinfo {year} {1996})}\BibitemShut {NoStop}%
\bibitem [{\citenamefont {Kim}(1997)}]{kim1997quasiprobability}%
  \BibitemOpen
  \bibfield  {author} {\bibinfo {author} {\bibfnamefont {M.~S.}\ \bibnamefont {Kim}},\ }\href {https://doi.org/10.1103/PhysRevA.56.3175} {\bibfield  {journal} {\bibinfo  {journal} {Physical Review A}\ }\textbf {\bibinfo {volume} {56}},\ \bibinfo {pages} {3175} (\bibinfo {year} {1997})}\BibitemShut {NoStop}%
\bibitem [{\citenamefont {Zambra}\ \emph {et~al.}(2005)\citenamefont {Zambra}, \citenamefont {Andreoni}, \citenamefont {Bondani}, \citenamefont {Gramegna}, \citenamefont {Genovese}, \citenamefont {Brida}, \citenamefont {Rossi},\ and\ \citenamefont {Paris}}]{Zambra2005}%
  \BibitemOpen
  \bibfield  {author} {\bibinfo {author} {\bibfnamefont {G.}~\bibnamefont {Zambra}}, \bibinfo {author} {\bibfnamefont {A.}~\bibnamefont {Andreoni}}, \bibinfo {author} {\bibfnamefont {M.}~\bibnamefont {Bondani}}, \bibinfo {author} {\bibfnamefont {M.}~\bibnamefont {Gramegna}}, \bibinfo {author} {\bibfnamefont {M.}~\bibnamefont {Genovese}}, \bibinfo {author} {\bibfnamefont {G.}~\bibnamefont {Brida}}, \bibinfo {author} {\bibfnamefont {A.}~\bibnamefont {Rossi}},\ and\ \bibinfo {author} {\bibfnamefont {M.~G.~A.}\ \bibnamefont {Paris}},\ }\href {https://doi.org/10.1103/PhysRevLett.95.063602} {\bibfield  {journal} {\bibinfo  {journal} {Physical Review Letters}\ }\textbf {\bibinfo {volume} {95}},\ \bibinfo {pages} {063602} (\bibinfo {year} {2005})}\BibitemShut {NoStop}%
\bibitem [{\citenamefont {Clark}\ \emph {et~al.}(2019)\citenamefont {Clark}, \citenamefont {Stokes},\ and\ \citenamefont {Beige}}]{Clark2019}%
  \BibitemOpen
  \bibfield  {author} {\bibinfo {author} {\bibfnamefont {L.~A.}\ \bibnamefont {Clark}}, \bibinfo {author} {\bibfnamefont {A.}~\bibnamefont {Stokes}},\ and\ \bibinfo {author} {\bibfnamefont {A.}~\bibnamefont {Beige}},\ }\href {https://doi.org/10.1103/PhysRevA.99.022102} {\bibfield  {journal} {\bibinfo  {journal} {Physical Review A}\ }\textbf {\bibinfo {volume} {99}},\ \bibinfo {pages} {022102} (\bibinfo {year} {2019})}\BibitemShut {NoStop}%
\bibitem [{\citenamefont {Wein}(2024)}]{Wein2024}%
  \BibitemOpen
  \bibfield  {author} {\bibinfo {author} {\bibfnamefont {S.~C.}\ \bibnamefont {Wein}},\ }\href {https://doi.org/10.1103/PhysRevA.109.023713} {\bibfield  {journal} {\bibinfo  {journal} {Physical Review A}\ }\textbf {\bibinfo {volume} {109}},\ \bibinfo {pages} {023713} (\bibinfo {year} {2024})}\BibitemShut {NoStop}%
\bibitem [{\citenamefont {Nunn}\ \emph {et~al.}(2021)\citenamefont {Nunn}, \citenamefont {Franson},\ and\ \citenamefont {Pittman}}]{Nunn2021}%
  \BibitemOpen
  \bibfield  {author} {\bibinfo {author} {\bibfnamefont {C.~M.}\ \bibnamefont {Nunn}}, \bibinfo {author} {\bibfnamefont {J.~D.}\ \bibnamefont {Franson}},\ and\ \bibinfo {author} {\bibfnamefont {T.~B.}\ \bibnamefont {Pittman}},\ }\href {https://doi.org/10.1103/PhysRevA.104.033717} {\bibfield  {journal} {\bibinfo  {journal} {Physical Review A}\ }\textbf {\bibinfo {volume} {104}},\ \bibinfo {pages} {033717} (\bibinfo {year} {2021})}\BibitemShut {NoStop}%
\bibitem [{\citenamefont {Nunn}\ \emph {et~al.}(2022)\citenamefont {Nunn}, \citenamefont {Franson},\ and\ \citenamefont {Pittman}}]{Nunn2022}%
  \BibitemOpen
  \bibfield  {author} {\bibinfo {author} {\bibfnamefont {C.~M.}\ \bibnamefont {Nunn}}, \bibinfo {author} {\bibfnamefont {J.~D.}\ \bibnamefont {Franson}},\ and\ \bibinfo {author} {\bibfnamefont {T.~B.}\ \bibnamefont {Pittman}},\ }\href {https://doi.org/10.1103/PhysRevA.105.033702} {\bibfield  {journal} {\bibinfo  {journal} {Physical Review A}\ }\textbf {\bibinfo {volume} {105}},\ \bibinfo {pages} {033702} (\bibinfo {year} {2022})}\BibitemShut {NoStop}%
\bibitem [{\citenamefont {B{\o}rkje}\ \emph {et~al.}(2011)\citenamefont {B{\o}rkje}, \citenamefont {Nunnenkamp},\ and\ \citenamefont {Girvin}}]{borkje2011proposal}%
  \BibitemOpen
  \bibfield  {author} {\bibinfo {author} {\bibfnamefont {K.}~\bibnamefont {B{\o}rkje}}, \bibinfo {author} {\bibfnamefont {A.}~\bibnamefont {Nunnenkamp}},\ and\ \bibinfo {author} {\bibfnamefont {S.}~\bibnamefont {Girvin}},\ }\href {https://doi.org/10.1103/PhysRevLett.107.123601} {\bibfield  {journal} {\bibinfo  {journal} {Physical Review Letters}\ }\textbf {\bibinfo {volume} {107}},\ \bibinfo {pages} {123601} (\bibinfo {year} {2011})}\BibitemShut {NoStop}%
\bibitem [{\citenamefont {Vanner}\ \emph {et~al.}(2013{\natexlab{b}})\citenamefont {Vanner}, \citenamefont {Aspelmeyer},\ and\ \citenamefont {Kim}}]{Vanner2013}%
  \BibitemOpen
  \bibfield  {author} {\bibinfo {author} {\bibfnamefont {M.~R.}\ \bibnamefont {Vanner}}, \bibinfo {author} {\bibfnamefont {M.}~\bibnamefont {Aspelmeyer}},\ and\ \bibinfo {author} {\bibfnamefont {M.~S.}\ \bibnamefont {Kim}},\ }\href {https://doi.org/10.1103/PhysRevLett.110.010504} {\bibfield  {journal} {\bibinfo  {journal} {Physical Review Letters}\ }\textbf {\bibinfo {volume} {110}},\ \bibinfo {pages} {010504} (\bibinfo {year} {2013}{\natexlab{b}})}\BibitemShut {NoStop}%
\bibitem [{\citenamefont {Lee}\ \emph {et~al.}(2012)\citenamefont {Lee}, \citenamefont {Sussman}, \citenamefont {Sprague}, \citenamefont {Michelberger}, \citenamefont {Reim}, \citenamefont {Nunn}, \citenamefont {Langford}, \citenamefont {Bustard}, \citenamefont {Jaksch},\ and\ \citenamefont {Walmsley}}]{Lee2012}%
  \BibitemOpen
  \bibfield  {author} {\bibinfo {author} {\bibfnamefont {K.}~\bibnamefont {Lee}}, \bibinfo {author} {\bibfnamefont {B.}~\bibnamefont {Sussman}}, \bibinfo {author} {\bibfnamefont {M.}~\bibnamefont {Sprague}}, \bibinfo {author} {\bibfnamefont {P.}~\bibnamefont {Michelberger}}, \bibinfo {author} {\bibfnamefont {K.}~\bibnamefont {Reim}}, \bibinfo {author} {\bibfnamefont {J.}~\bibnamefont {Nunn}}, \bibinfo {author} {\bibfnamefont {N.}~\bibnamefont {Langford}}, \bibinfo {author} {\bibfnamefont {P.}~\bibnamefont {Bustard}}, \bibinfo {author} {\bibfnamefont {D.}~\bibnamefont {Jaksch}},\ and\ \bibinfo {author} {\bibfnamefont {I.}~\bibnamefont {Walmsley}},\ }\href {https://doi.org/10.1038/nphoton.2011.296} {\bibfield  {journal} {\bibinfo  {journal} {Nature Photonics}\ }\textbf {\bibinfo {volume} {6}},\ \bibinfo {pages} {41} (\bibinfo {year} {2012})}\BibitemShut {NoStop}%
\bibitem [{\citenamefont {Riedinger}\ \emph {et~al.}(2016)\citenamefont {Riedinger}, \citenamefont {Hong}, \citenamefont {Norte}, \citenamefont {Slater}, \citenamefont {Shang}, \citenamefont {Krause}, \citenamefont {Anant}, \citenamefont {Aspelmeyer},\ and\ \citenamefont {Gr{\"o}blacher}}]{riedinger2016non}%
  \BibitemOpen
  \bibfield  {author} {\bibinfo {author} {\bibfnamefont {R.}~\bibnamefont {Riedinger}}, \bibinfo {author} {\bibfnamefont {S.}~\bibnamefont {Hong}}, \bibinfo {author} {\bibfnamefont {R.~A.}\ \bibnamefont {Norte}}, \bibinfo {author} {\bibfnamefont {J.~A.}\ \bibnamefont {Slater}}, \bibinfo {author} {\bibfnamefont {J.}~\bibnamefont {Shang}}, \bibinfo {author} {\bibfnamefont {A.~G.}\ \bibnamefont {Krause}}, \bibinfo {author} {\bibfnamefont {V.}~\bibnamefont {Anant}}, \bibinfo {author} {\bibfnamefont {M.}~\bibnamefont {Aspelmeyer}},\ and\ \bibinfo {author} {\bibfnamefont {S.}~\bibnamefont {Gr{\"o}blacher}},\ }\href {https://doi.org/10.1038/nature16536} {\bibfield  {journal} {\bibinfo  {journal} {Nature}\ }\textbf {\bibinfo {volume} {530}},\ \bibinfo {pages} {313} (\bibinfo {year} {2016})}\BibitemShut {NoStop}%
\bibitem [{\citenamefont {Enzian}\ \emph {et~al.}(2021{\natexlab{a}})\citenamefont {Enzian}, \citenamefont {Price}, \citenamefont {Freisem}, \citenamefont {Nunn}, \citenamefont {Janousek}, \citenamefont {Buchler}, \citenamefont {Lam},\ and\ \citenamefont {Vanner}}]{Enzian2021}%
  \BibitemOpen
  \bibfield  {author} {\bibinfo {author} {\bibfnamefont {G.}~\bibnamefont {Enzian}}, \bibinfo {author} {\bibfnamefont {J.~J.}\ \bibnamefont {Price}}, \bibinfo {author} {\bibfnamefont {L.}~\bibnamefont {Freisem}}, \bibinfo {author} {\bibfnamefont {J.}~\bibnamefont {Nunn}}, \bibinfo {author} {\bibfnamefont {J.}~\bibnamefont {Janousek}}, \bibinfo {author} {\bibfnamefont {B.~C.}\ \bibnamefont {Buchler}}, \bibinfo {author} {\bibfnamefont {P.~K.}\ \bibnamefont {Lam}},\ and\ \bibinfo {author} {\bibfnamefont {M.~R.}\ \bibnamefont {Vanner}},\ }\href {https://doi.org/10.1103/PhysRevLett.126.033601} {\bibfield  {journal} {\bibinfo  {journal} {Physical Review Letters}\ }\textbf {\bibinfo {volume} {126}},\ \bibinfo {pages} {33601} (\bibinfo {year} {2021}{\natexlab{a}})}\BibitemShut {NoStop}%
\bibitem [{\citenamefont {Enzian}\ \emph {et~al.}(2021{\natexlab{b}})\citenamefont {Enzian}, \citenamefont {Freisem}, \citenamefont {Price}, \citenamefont {Svela}, \citenamefont {Clarke}, \citenamefont {Shajilal}, \citenamefont {Janousek}, \citenamefont {Buchler}, \citenamefont {Lam},\ and\ \citenamefont {Vanner}}]{Enzian2021_2}%
  \BibitemOpen
  \bibfield  {author} {\bibinfo {author} {\bibfnamefont {G.}~\bibnamefont {Enzian}}, \bibinfo {author} {\bibfnamefont {L.}~\bibnamefont {Freisem}}, \bibinfo {author} {\bibfnamefont {J.~J.}\ \bibnamefont {Price}}, \bibinfo {author} {\bibfnamefont {A.}~\bibnamefont {Svela}}, \bibinfo {author} {\bibfnamefont {J.}~\bibnamefont {Clarke}}, \bibinfo {author} {\bibfnamefont {B.}~\bibnamefont {Shajilal}}, \bibinfo {author} {\bibfnamefont {J.}~\bibnamefont {Janousek}}, \bibinfo {author} {\bibfnamefont {B.~C.}\ \bibnamefont {Buchler}}, \bibinfo {author} {\bibfnamefont {P.~K.}\ \bibnamefont {Lam}},\ and\ \bibinfo {author} {\bibfnamefont {M.~R.}\ \bibnamefont {Vanner}},\ }\href {https://doi.org/10.1103/PhysRevLett.127.243601} {\bibfield  {journal} {\bibinfo  {journal} {Physical Review Letters}\ }\textbf {\bibinfo {volume} {127}},\ \bibinfo {pages} {243601} (\bibinfo {year} {2021}{\natexlab{b}})}\BibitemShut {NoStop}%
\bibitem [{\citenamefont {Patel}\ \emph {et~al.}(2021)\citenamefont {Patel}, \citenamefont {McKenna}, \citenamefont {Wang}, \citenamefont {Witmer}, \citenamefont {Jiang}, \citenamefont {{Van Laer}}, \citenamefont {Sarabalis},\ and\ \citenamefont {Safavi-Naeini}}]{Patel2021}%
  \BibitemOpen
  \bibfield  {author} {\bibinfo {author} {\bibfnamefont {R.~N.}\ \bibnamefont {Patel}}, \bibinfo {author} {\bibfnamefont {T.~P.}\ \bibnamefont {McKenna}}, \bibinfo {author} {\bibfnamefont {Z.}~\bibnamefont {Wang}}, \bibinfo {author} {\bibfnamefont {J.~D.}\ \bibnamefont {Witmer}}, \bibinfo {author} {\bibfnamefont {W.}~\bibnamefont {Jiang}}, \bibinfo {author} {\bibfnamefont {R.}~\bibnamefont {{Van Laer}}}, \bibinfo {author} {\bibfnamefont {C.~J.}\ \bibnamefont {Sarabalis}},\ and\ \bibinfo {author} {\bibfnamefont {A.~H.}\ \bibnamefont {Safavi-Naeini}},\ }\href {https://doi.org/10.1103/PhysRevLett.127.133602} {\bibfield  {journal} {\bibinfo  {journal} {Physical Review Letters}\ }\textbf {\bibinfo {volume} {127}},\ \bibinfo {pages} {133602} (\bibinfo {year} {2021})}\BibitemShut {NoStop}%
\bibitem [{\citenamefont {Patil}\ \emph {et~al.}(2022)\citenamefont {Patil}, \citenamefont {Yu}, \citenamefont {Frazier}, \citenamefont {Wang}, \citenamefont {Johnson}, \citenamefont {Fox}, \citenamefont {Reichel},\ and\ \citenamefont {Harris}}]{Patil2022}%
  \BibitemOpen
  \bibfield  {author} {\bibinfo {author} {\bibfnamefont {Y.~S.}\ \bibnamefont {Patil}}, \bibinfo {author} {\bibfnamefont {J.}~\bibnamefont {Yu}}, \bibinfo {author} {\bibfnamefont {S.}~\bibnamefont {Frazier}}, \bibinfo {author} {\bibfnamefont {Y.}~\bibnamefont {Wang}}, \bibinfo {author} {\bibfnamefont {K.}~\bibnamefont {Johnson}}, \bibinfo {author} {\bibfnamefont {J.}~\bibnamefont {Fox}}, \bibinfo {author} {\bibfnamefont {J.}~\bibnamefont {Reichel}},\ and\ \bibinfo {author} {\bibfnamefont {J.~G.}\ \bibnamefont {Harris}},\ }\href {https://doi.org/10.1103/PhysRevLett.128.183601} {\bibfield  {journal} {\bibinfo  {journal} {Physical Review Letters}\ }\textbf {\bibinfo {volume} {128}},\ \bibinfo {pages} {183601} (\bibinfo {year} {2022})}\BibitemShut {NoStop}%
\bibitem [{\citenamefont {Qiu}\ \emph {et~al.}(2020)\citenamefont {Qiu}, \citenamefont {Shomroni}, \citenamefont {Seidler},\ and\ \citenamefont {Kippenberg}}]{Qiu2020}%
  \BibitemOpen
  \bibfield  {author} {\bibinfo {author} {\bibfnamefont {L.}~\bibnamefont {Qiu}}, \bibinfo {author} {\bibfnamefont {I.}~\bibnamefont {Shomroni}}, \bibinfo {author} {\bibfnamefont {P.}~\bibnamefont {Seidler}},\ and\ \bibinfo {author} {\bibfnamefont {T.~J.}\ \bibnamefont {Kippenberg}},\ }\href {https://doi.org/10.1103/PhysRevLett.124.173601} {\bibfield  {journal} {\bibinfo  {journal} {Physical Review Letters}\ }\textbf {\bibinfo {volume} {124}},\ \bibinfo {pages} {173601} (\bibinfo {year} {2020})}\BibitemShut {NoStop}%
\bibitem [{\citenamefont {Peterson}\ \emph {et~al.}(2016)\citenamefont {Peterson}, \citenamefont {Purdy}, \citenamefont {Kampel}, \citenamefont {Andrews}, \citenamefont {Yu}, \citenamefont {Lehnert},\ and\ \citenamefont {Regal}}]{Peterson2016}%
  \BibitemOpen
  \bibfield  {author} {\bibinfo {author} {\bibfnamefont {R.}~\bibnamefont {Peterson}}, \bibinfo {author} {\bibfnamefont {T.}~\bibnamefont {Purdy}}, \bibinfo {author} {\bibfnamefont {N.}~\bibnamefont {Kampel}}, \bibinfo {author} {\bibfnamefont {R.}~\bibnamefont {Andrews}}, \bibinfo {author} {\bibfnamefont {P.-L.}\ \bibnamefont {Yu}}, \bibinfo {author} {\bibfnamefont {K.}~\bibnamefont {Lehnert}},\ and\ \bibinfo {author} {\bibfnamefont {C.}~\bibnamefont {Regal}},\ }\href {http://link.aps.org/doi/10.1103/PhysRevLett.116.063601} {\bibfield  {journal} {\bibinfo  {journal} {Physical Review Letters}\ }\textbf {\bibinfo {volume} {116}},\ \bibinfo {pages} {063601} (\bibinfo {year} {2016})}\BibitemShut {NoStop}%
\bibitem [{\citenamefont {Bl\'azquez~Mart\'{\i}nez}\ \emph {et~al.}(2024)\citenamefont {Bl\'azquez~Mart\'{\i}nez}, \citenamefont {Wiedemann}, \citenamefont {Zhu}, \citenamefont {Geilen},\ and\ \citenamefont {Stiller}}]{Blazquez2024}%
  \BibitemOpen
  \bibfield  {author} {\bibinfo {author} {\bibfnamefont {L.}~\bibnamefont {Bl\'azquez~Mart\'{\i}nez}}, \bibinfo {author} {\bibfnamefont {P.}~\bibnamefont {Wiedemann}}, \bibinfo {author} {\bibfnamefont {C.}~\bibnamefont {Zhu}}, \bibinfo {author} {\bibfnamefont {A.}~\bibnamefont {Geilen}},\ and\ \bibinfo {author} {\bibfnamefont {B.}~\bibnamefont {Stiller}},\ }\href {https://doi.org/10.1103/PhysRevLett.132.023603} {\bibfield  {journal} {\bibinfo  {journal} {Physical Review Letters}\ }\textbf {\bibinfo {volume} {132}},\ \bibinfo {pages} {023603} (\bibinfo {year} {2024})}\BibitemShut {NoStop}%
\bibitem [{\citenamefont {Clarke}\ \emph {et~al.}(2024)\citenamefont {Clarke}, \citenamefont {Cryer-Jenkins}, \citenamefont {Gupta}, \citenamefont {Major}, \citenamefont {Zhang}, \citenamefont {Enzian}, \citenamefont {Szczykulska}, \citenamefont {Leung}, \citenamefont {Rathee}, \citenamefont {Tan}, \citenamefont {Svela}, \citenamefont {Beige}, \citenamefont {M{\o}lmer},\ and\ \citenamefont {Vanner}}]{cryer2023mechanical}%
  \BibitemOpen
  \bibfield  {author} {\bibinfo {author} {\bibfnamefont {J.}~\bibnamefont {Clarke}}, \bibinfo {author} {\bibfnamefont {E.~A.}\ \bibnamefont {Cryer-Jenkins}}, \bibinfo {author} {\bibfnamefont {A.}~\bibnamefont {Gupta}}, \bibinfo {author} {\bibfnamefont {K.~D.}\ \bibnamefont {Major}}, \bibinfo {author} {\bibfnamefont {J.}~\bibnamefont {Zhang}}, \bibinfo {author} {\bibfnamefont {G.}~\bibnamefont {Enzian}}, \bibinfo {author} {\bibfnamefont {M.}~\bibnamefont {Szczykulska}}, \bibinfo {author} {\bibfnamefont {A.~C.}\ \bibnamefont {Leung}}, \bibinfo {author} {\bibfnamefont {H.}~\bibnamefont {Rathee}}, \bibinfo {author} {\bibfnamefont {A.~K.~C.}\ \bibnamefont {Tan}}, \bibinfo {author} {\bibfnamefont {A.~{\O}.}\ \bibnamefont {Svela}}, \bibinfo {author} {\bibfnamefont {A.}~\bibnamefont {Beige}}, \bibinfo {author} {\bibfnamefont {K.}~\bibnamefont {M{\o}lmer}},\ and\ \bibinfo {author} {\bibfnamefont {M.~R.}\ \bibnamefont {Vanner}},\ }\href {https://arxiv.org/abs/2408.01735} {\bibfield  {journal} {\bibinfo  {journal}
  {arXiv:2408.01735}\ } (\bibinfo {year} {2024})}\BibitemShut {NoStop}%
\bibitem [{\citenamefont {Boyd}(1992)}]{Boyd1992}%
  \BibitemOpen
  \bibfield  {author} {\bibinfo {author} {\bibfnamefont {R.}~\bibnamefont {Boyd}},\ }\href@noop {} {\emph {\bibinfo {title} {{Nonlinear Optics}}}}\ (\bibinfo  {publisher} {San Diego, CA (United States); Academic Press Inc.},\ \bibinfo {year} {1992})\BibitemShut {NoStop}%
\bibitem [{\citenamefont {Enzian}\ \emph {et~al.}(2019)\citenamefont {Enzian}, \citenamefont {Szczykulska}, \citenamefont {Silver}, \citenamefont {Del~Bino}, \citenamefont {Zhang}, \citenamefont {Walmsley}, \citenamefont {Del’Haye},\ and\ \citenamefont {Vanner}}]{Enzian2019}%
  \BibitemOpen
  \bibfield  {author} {\bibinfo {author} {\bibfnamefont {G.}~\bibnamefont {Enzian}}, \bibinfo {author} {\bibfnamefont {M.}~\bibnamefont {Szczykulska}}, \bibinfo {author} {\bibfnamefont {J.}~\bibnamefont {Silver}}, \bibinfo {author} {\bibfnamefont {L.}~\bibnamefont {Del~Bino}}, \bibinfo {author} {\bibfnamefont {S.}~\bibnamefont {Zhang}}, \bibinfo {author} {\bibfnamefont {I.~A.}\ \bibnamefont {Walmsley}}, \bibinfo {author} {\bibfnamefont {P.}~\bibnamefont {Del’Haye}},\ and\ \bibinfo {author} {\bibfnamefont {M.~R.}\ \bibnamefont {Vanner}},\ }\href {https://opg.optica.org/optica/fulltext.cfm?uri=optica-6-1-7&id=403306} {\bibfield  {journal} {\bibinfo  {journal} {Optica}\ }\textbf {\bibinfo {volume} {6}},\ \bibinfo {pages} {7} (\bibinfo {year} {2019})}\BibitemShut {NoStop}%
\bibitem [{sup()}]{supp}%
  \BibitemOpen
  \href@noop {} {}\bibinfo {note} {See the Supplemental Material for further experimental and theoretical details.}\BibitemShut {Stop}%
\bibitem [{\citenamefont {Kristensen}\ \emph {et~al.}(2024)\citenamefont {Kristensen}, \citenamefont {Kralj}, \citenamefont {Langman},\ and\ \citenamefont {Schliesser}}]{Kristensen2024}%
  \BibitemOpen
  \bibfield  {author} {\bibinfo {author} {\bibfnamefont {M.~B.}\ \bibnamefont {Kristensen}}, \bibinfo {author} {\bibfnamefont {N.}~\bibnamefont {Kralj}}, \bibinfo {author} {\bibfnamefont {E.~C.}\ \bibnamefont {Langman}},\ and\ \bibinfo {author} {\bibfnamefont {A.}~\bibnamefont {Schliesser}},\ }\href {https://doi.org/10.1103/PhysRevLett.132.100802} {\bibfield  {journal} {\bibinfo  {journal} {Physical Review Letters}\ }\textbf {\bibinfo {volume} {132}},\ \bibinfo {pages} {100802} (\bibinfo {year} {2024})}\BibitemShut {NoStop}%
\bibitem [{\citenamefont {Bahl}\ \emph {et~al.}(2012)\citenamefont {Bahl}, \citenamefont {Tomes}, \citenamefont {Marquardt},\ and\ \citenamefont {Carmon}}]{Bahl2012}%
  \BibitemOpen
  \bibfield  {author} {\bibinfo {author} {\bibfnamefont {G.}~\bibnamefont {Bahl}}, \bibinfo {author} {\bibfnamefont {M.}~\bibnamefont {Tomes}}, \bibinfo {author} {\bibfnamefont {F.}~\bibnamefont {Marquardt}},\ and\ \bibinfo {author} {\bibfnamefont {T.}~\bibnamefont {Carmon}},\ }\href {https://www.nature.com/articles/nphys2206} {\bibfield  {journal} {\bibinfo  {journal} {Nature Physics}\ }\textbf {\bibinfo {volume} {8}},\ \bibinfo {pages} {203} (\bibinfo {year} {2012})}\BibitemShut {NoStop}%
\bibitem [{\citenamefont {Verhagen}\ \emph {et~al.}(2012)\citenamefont {Verhagen}, \citenamefont {Deléglise}, \citenamefont {Weis}, \citenamefont {Schliesser},\ and\ \citenamefont {Kippenberg}}]{Verhagen2012}%
  \BibitemOpen
  \bibfield  {author} {\bibinfo {author} {\bibfnamefont {E.}~\bibnamefont {Verhagen}}, \bibinfo {author} {\bibfnamefont {S.}~\bibnamefont {Deléglise}}, \bibinfo {author} {\bibfnamefont {S.}~\bibnamefont {Weis}}, \bibinfo {author} {\bibfnamefont {A.}~\bibnamefont {Schliesser}},\ and\ \bibinfo {author} {\bibfnamefont {T.~J.}\ \bibnamefont {Kippenberg}},\ }\href {http://www.nature.com/nature/journal/v482/n7383/full/nature10787.html} {\bibfield  {journal} {\bibinfo  {journal} {Nature}\ }\textbf {\bibinfo {volume} {482}},\ \bibinfo {pages} {63} (\bibinfo {year} {2012})}\BibitemShut {NoStop}%
\bibitem [{\citenamefont {Cohen}\ \emph {et~al.}(2015)\citenamefont {Cohen}, \citenamefont {Meenehan}, \citenamefont {MacCabe}, \citenamefont {Gr{\"o}blacher}, \citenamefont {Safavi-Naeini}, \citenamefont {Marsili}, \citenamefont {Shaw},\ and\ \citenamefont {Painter}}]{cohen2015phonon}%
  \BibitemOpen
  \bibfield  {author} {\bibinfo {author} {\bibfnamefont {J.~D.}\ \bibnamefont {Cohen}}, \bibinfo {author} {\bibfnamefont {S.~M.}\ \bibnamefont {Meenehan}}, \bibinfo {author} {\bibfnamefont {G.~S.}\ \bibnamefont {MacCabe}}, \bibinfo {author} {\bibfnamefont {S.}~\bibnamefont {Gr{\"o}blacher}}, \bibinfo {author} {\bibfnamefont {A.~H.}\ \bibnamefont {Safavi-Naeini}}, \bibinfo {author} {\bibfnamefont {F.}~\bibnamefont {Marsili}}, \bibinfo {author} {\bibfnamefont {M.~D.}\ \bibnamefont {Shaw}},\ and\ \bibinfo {author} {\bibfnamefont {O.}~\bibnamefont {Painter}},\ }\href {https://www.nature.com/articles/nature14349} {\bibfield  {journal} {\bibinfo  {journal} {Nature}\ }\textbf {\bibinfo {volume} {520}},\ \bibinfo {pages} {522} (\bibinfo {year} {2015})}\BibitemShut {NoStop}%
\bibitem [{\citenamefont {Cryer-Jenkins}\ \emph {et~al.}(2023)\citenamefont {Cryer-Jenkins}, \citenamefont {Enzian}, \citenamefont {Freisem}, \citenamefont {Moroney}, \citenamefont {Price}, \citenamefont {Svela}, \citenamefont {Major},\ and\ \citenamefont {Vanner}}]{cryer2023second}%
  \BibitemOpen
  \bibfield  {author} {\bibinfo {author} {\bibfnamefont {E.~A.}\ \bibnamefont {Cryer-Jenkins}}, \bibinfo {author} {\bibfnamefont {G.}~\bibnamefont {Enzian}}, \bibinfo {author} {\bibfnamefont {L.}~\bibnamefont {Freisem}}, \bibinfo {author} {\bibfnamefont {N.}~\bibnamefont {Moroney}}, \bibinfo {author} {\bibfnamefont {J.~J.}\ \bibnamefont {Price}}, \bibinfo {author} {\bibfnamefont {A.~{\O}.}\ \bibnamefont {Svela}}, \bibinfo {author} {\bibfnamefont {K.~D.}\ \bibnamefont {Major}},\ and\ \bibinfo {author} {\bibfnamefont {M.~R.}\ \bibnamefont {Vanner}},\ }\href {https://opg.optica.org/viewmedia.cfm?uri=optica-10-11-1432&seq=0&html=true} {\bibfield  {journal} {\bibinfo  {journal} {Optica}\ }\textbf {\bibinfo {volume} {10}},\ \bibinfo {pages} {1432} (\bibinfo {year} {2023})}\BibitemShut {NoStop}%
\bibitem [{\citenamefont {Milburn}\ \emph {et~al.}(2016)\citenamefont {Milburn}, \citenamefont {Kim},\ and\ \citenamefont {Vanner}}]{Milburn2016}%
  \BibitemOpen
  \bibfield  {author} {\bibinfo {author} {\bibfnamefont {T.~J.}\ \bibnamefont {Milburn}}, \bibinfo {author} {\bibfnamefont {M.~S.}\ \bibnamefont {Kim}},\ and\ \bibinfo {author} {\bibfnamefont {M.~R.}\ \bibnamefont {Vanner}},\ }\href {https://doi.org/10.1103/PhysRevA.93.053818} {\bibfield  {journal} {\bibinfo  {journal} {Physical Review A}\ }\textbf {\bibinfo {volume} {93}},\ \bibinfo {pages} {053818} (\bibinfo {year} {2016})}\BibitemShut {NoStop}%
\bibitem [{\citenamefont {Galinskiy}\ \emph {et~al.}(2023)\citenamefont {Galinskiy}, \citenamefont {Enzian}, \citenamefont {Parniak},\ and\ \citenamefont {Polzik}}]{galinskiy2023non}%
  \BibitemOpen
  \bibfield  {author} {\bibinfo {author} {\bibfnamefont {I.}~\bibnamefont {Galinskiy}}, \bibinfo {author} {\bibfnamefont {G.}~\bibnamefont {Enzian}}, \bibinfo {author} {\bibfnamefont {M.}~\bibnamefont {Parniak}},\ and\ \bibinfo {author} {\bibfnamefont {E.}~\bibnamefont {Polzik}},\ }\href {https://doi.org/10.48550/arXiv.2312.05641} {\bibfield  {journal} {\bibinfo  {journal} {arXiv:2312.05641}\ } (\bibinfo {year} {2023})}\BibitemShut {NoStop}%
\end{thebibliography}
\providecommand{\noopsort}[1]{}\providecommand{\singleletter}[1]{#1}%
%


\onecolumngrid
\clearpage
\newgeometry{left=1.5cm,right=1.5cm,top=1.5cm,bottom=1.5cm}
\setcounter{equation}{0}
\def\theequation{S\arabic{equation}}
\setcounter{figure}{0}
\def\thefigure{S\arabic{figure}}
\pagenumbering{roman}

\begin{center}
\textbf{\large{Something from Nothing: \\ Enhanced Laser Cooling of a Mechanical Resonator via Zero-Photon Detection\\Supplemental Material}}
\end{center}
\vspace{10pt}

\begin{centering}
Evan A.~Cryer-Jenkins,$^{1,*}$~
Kyle D.~Major,$^{1,*}$~
Jack Clarke,$^1$~
Georg Enzian,$^{1,2}$\\
Magdalena Szczykulska,$^4$~
Jinglei Zhang\,,$^{2,3}$~
Arjun Gupta,$^1$~
Anthony C.~Leung,$^1$~
Harsh Rathee,$^1$~
Andreas {\O}.~Svela,$^{1}$~
Anthony K.~C.~Tan,$^1$~
Almut Beige,$^5$~
Klaus M{\o}lmer,$^6$~
and Michael R.~Vanner$^{1}$\\
\end{centering}

\vspace{16pt}

\begin{centering}
\textit{\small
$^*$These authors contributed equally to this work and are listed alphabetically.\\
$^1${Quantum Measurement Lab, Blackett Laboratory, Imperial College London, London SW7 2BW, United Kingdom}\\
$^2${Clarendon Laboratory, Department of Physics, University of Oxford, OX1 3PU, United Kingdom}\\
$^3${Institute for Quantum Computing, University of Waterloo, Waterloo, Ontario, N2L 3G1, Canada}\\
$^4${Department of Physics \& Astronomy, University of Waterloo, Waterloo, Ontario, N2L 3G1, Canada}\\
$^5${The School of Physics and Astronomy, University of Leeds, Leeds LS2 9JT, United Kingdom}\\
$^6${Niels Bohr Institute, University of Copenhagen, Blegdamsvej 17, 2100 Copenhagen, Denmark}\\
}
\end{centering}

\begin{quote}
{\small{In this Supplemental, we provide further details of our theoretical model and experimental apparatus. We first mathematically discuss the dynamics of the enhanced cooling via zero-photon measurement. Secondly, we describe the regime where an adiabatic approximation can be made to the anti-Stokes cavity mode and the output optical mode becomes an ideal proxy for the state of the mechanical oscillator. Thirdly, we discuss the limits of mechanical cooling via zero-photon detection analytically. After this, we discuss the heterodyne signal normalization, and, finally, we give a more in-depth discussion of our experimental system characterization and provide a table of the relevant system parameters. 
}}
\end{quote}

\onecolumngrid

\subsubsection{Dynamics and cooling rates of zero-photon-detection-enhanced laser cooling}
The equations that describe continuous cooling via zero-photon measurement from our companion theory paper~[\href{https://arxiv.org/abs/2408.01735}{J. Clarke, E. A. Cryer-Jenkins, A. Gupta, K. D. Major, \emph{et al.,} arXiv:2408.01735 (2024)}] are:
\begin{align}
    &\rmi\dfrac{d}{\rmd{t}}\expval{\ad b}=G\left[\expval{\ad a}-\expval{\bd b}\right]-\rmi\left(\kappa+\gamma\right)\expval{\ad b}-2\rmi{\eta}\kappa_{ex}\expval{\ad a}\expval{\ad b},\label{eq:zc_as_cde_1}\\
    &\rmi\dfrac{d}{\rmd{t}}\expval{a\bd}=-G\left[\expval{\ad a}-\expval{\bd b}\right]-\rmi\left(\kappa+\gamma\right)\expval{a\bd}-2\rmi{\eta}\kappa_{ex}\expval{\ad a}\expval{a \bd},\label{eq:zc_as_cde_2}\\
    &\dfrac{d}{\rmd{t}}\expval{\ad a}=-\rmi G \expval{\ad b-a\bd}-2\kappa\expval{\ad a}-2{\eta}\kappa_{ex}\expval{\ad a}^2,\label{eq:zc_as_cde_3}\\
    &\dfrac{d}{\rmd{t}}\expval{\bd b}=\rmi G \expval{\ad b-a\bd}-2\gamma\expval{\bd b}+2\gamma\bar{N}-2{\eta}\kappa_{ex}\expval{a\bd}\expval{\ad b}.\label{eq:zc_as_cde_4}
\end{align}
Here, we follow the notation introduced in the main text and we have also used the phase symmetry of the optical and mechanical states to simplify these expressions. Moreover, for each $\rmd{t}$, $\mathcal{P}_{0}=1-2\eta\kappa_{ex}\expval{\ad a}\rmd{t}$ is the probability for a zero-photon detection and $\mathcal{P}_{1}=2\eta\kappa_{ex}\expval{\ad a}\rmd{t}$ is the probability for a single-photon detection. The total probability for a particular record of outcomes is then found by multiplying these corresponding probabilities at every instant in time. Thus, the probability to observe a string of zero-photon detection events of duration $T$ is given by the product $\Pi_{m=0}^{N}\left[1-2\eta\kappa_{ex}\expval{(\ad a)(t_m)}\rmd{t}\right]$, where $T=N\rmd{t}$, $m=0,1,\ldots,N$, and $t_m=m\rmd{t}$.

Eq.~\eqref{eq:zc_as_cde_4} may be used to calculate the rate of mechanical cooling via zero-photon detection. In particular, after the steady-state of laser cooling has been reached, the rate of cooling at the onset of zero-photon detection is
\begin{eqnarray}
    \dfrac{\rmd\expval{\bd b}}{\rmd t}&=&-\dfrac{2G^2\bar{N}\gamma^2\kappa_{ex}\kappa^2\eta}{(\gamma+\kappa)^2(G^2+\gamma\kappa)^2},\label{eq:supp_zero_click_onset}    
\end{eqnarray}
which is $-0.32\,\times 10^9\,\mathrm{s}^{-1}$, for $\eta=0.32\%$ and the values listed in Table~\ref{table:parameters} below. We also note that Eq.~\eqref{eq:supp_zero_click_onset} is proportional to $\eta$, which explains the linear scaling of the zero-photon signal with $\eta$ in Fig.~\hyperref[fig:fig2]{2(b)} of the main text for the single time-step measurement.

The steady state of enhanced cooling via continuous zero-photon detection is obtained by setting Eqs~\eqref{eq:zc_as_cde_1} to \eqref{eq:zc_as_cde_4} equal to zero.  In this case, the rates associated with rethermalization, laser cooling, and cooling via zero-photon detection balance. For $\eta=0.32\%$ and the values listed in Table~\ref{table:parameters}, the rates in the steady state of continuous zero-photon detection may be identified from the right-hand-side of Eq.~\eqref{eq:zc_as_cde_4} as:
\begin{itemize}
    \item Rethermalization rate: $2\gamma\left(\bar{N}-\expval{\bd b}\right)=1.97\,\times 10^9\,\mathrm{s}^{-1}$,
    \item Laser-cooling rate: $\rmi G \expval{\ad b-a\bd}=-1.65\,\times 10^9\,\mathrm{s}^{-1}$,
    \item Rate of cooling via zero-photon detection: $-2{\eta}\kappa_{ex}\expval{a\bd}\expval{\ad b}=-0.32\,\times 10^9\,\mathrm{s}^{-1}$.
\end{itemize}
Note that although $\kappa_{ex}$ and $\kappa$ do not appear explicitly in the rethermilzation and laser-cooling rates, the dependence on $\kappa_{ex}$ and $\kappa$ enter via the steady-state values of $\expval{\bd b}$, $\expval{\ad b}$, and $\expval{a\bd}$.

\subsubsection{Adiabatic approximation}
In the regime where the external amplitude coupling rate $\kappa_{ex}$ dominates over all other optical and mechanical rates, the cavity field may be adiabatically eliminated, i.e. $\dot{a}\simeq0$. Thus, the optical field at the detector closely follows the mechanical mode, which enables the limit of mechanical cooling via zero-photon detection to be derived. Furthermore, to derive this limit, we consider the scenario where the cavity is dominated by external coupling $\kappa\approx\kappa_{ex}$. This condition, combined with the efficiency outside the cavity of $\eta=1$, ensures that the full signal reaches the photon-counting detector to give the best zero-photon-detection enhanced cooling.

By taking $\dot{a}\simeq0$ and $\kappa\approx\kappa_{ex}$ in Eqs~\eqref{eq:zc_as_cde_1} to \eqref{eq:zc_as_cde_4}, one may arrive at the following equation of motion for the mechanical occupation: 
\begin{eqnarray}
    \dfrac{\rmd\expval{\bd b}}{\rmd t}&=&-\eta\kl{\expval{\bd b}}^2-\left(\kl+2\gamma\right)\expval{\bd b}+2\gamma\bar{N}.\label{eq:supp_zero_click_EOM}
\end{eqnarray}
Here, $\kl=2G^2/\kappa_{ex}$ is the optomechanical measurement rate. The solution to Eq.~\eqref{eq:supp_zero_click_EOM} is given by
\begin{eqnarray}
    \expval{\left(\bd b\right)(t)}&=&\frac{1}{2\eta\kl}\left\{-(2\gamma+\kl)+\xi\left[\dfrac{\nu+\xi\tanh{\frac{1}{2}\xi t}}{\xi+\nu\tanh{\frac{1}{2}\xi t}}\right]\right\},\label{eq:supp_zero_click_with_t}\\
    \xi&=&\sqrt{(2\gamma+\kl)^2+8\eta\kl\gamma\bar{N}},\\
    \nu&=&\left[2\eta\expval{\left(\bd b\right)(0)}+1\right]\kl+2\gamma,    
\end{eqnarray}
with the initial condition $\expval{\left(\bd b\right)(0)}$. For an adiabatic cavity, the limit of laser cooling corresponds to $\xi t\gg1$ and $\eta=0$ in Eq.~\eqref{eq:supp_zero_click_with_t}, which gives a mechanical occupation of
\begin{eqnarray}
\bar{n}_{LC}=\dfrac{\bar{N}}{1+C},\label{eq:supp_lasercool_inf}
\end{eqnarray}
where $C=\kl/2\gamma=G^2/\kappa_{ex}\gamma$ is the optomechanical cooperativity. The limit of zero-photon-detection-enhanced laser cooling is achieved for a long string of zero-photon detection events corresponding to $\xi t\gg1$ in Eq.~\eqref{eq:supp_zero_click_with_t} with $0<\eta\leq1$, giving a mechanical occupation of
\begin{eqnarray} 
 \bar{n}_{lim} =\dfrac{-(1+C)+\sqrt{(1+C)^2+4\eta\bar{N} C}}{2\eta C}.\label{eq:supp_zerocool_inf} 
\end{eqnarray} 
By using Eqs~\eqref{eq:supp_lasercool_inf} and \eqref{eq:supp_zerocool_inf}, one can show
the limit of zero-photon-measurement-enhanced laser cooling $\bar{n}_{lim}$ is below a target value $\bar{n}_{*}$ when
\begin{equation}
    \bar{n}_{LC}<\dfrac{\eta C \bar{n}_{*}^{2}+\bar{n}_{*}(1+C)}{1+C}.\label{seq:condition}
\end{equation}
The right-hand side of Eq.~\eqref{seq:condition} is maximised for $\eta=1$ and $C\rightarrow\infty$, which gives $\bar{n}_{LC}<\bar{n}_{*}^{2}+\bar{n}_{*}$. Hence, to reduce the mean thermal occupation below $1$ using zero-photon detection, one requires that the laser-cooled state is cooled to at least $\bar{n}_{LC}=2$.

The rate of change of the mechanical occupation due to zero-photon detection can be calculated via Eq.~\eqref{eq:supp_zero_click_EOM}. For a mechanical state initially in thermal equilibrium with its environment, at the onset of zero-photon-enhanced laser cooling this rate of change is given by 
\begin{equation}
    \dfrac{\rmd\expval{\bd b}}{\rmd t}=-(1+\eta\bar{N})\bar{N}\kl.
\end{equation}
Here, the contribution from laser cooling can be observed by setting $\eta=0$. Furthermore, once the laser-cooled steady state has been reached, the rate of change for the mechanical occupation at the onset of enhanced cooling via zero-photon detection is
\begin{eqnarray}
    \dfrac{\rmd\expval{\bd b}}{\rmd t}&=&-\dfrac{ \eta\bar{N}^2\kl}{(1+C)^2}\label{eq:supp_zero_click_onset_ad},
\end{eqnarray}
which agrees with Eq.~\eqref{eq:supp_zero_click_onset} when $\kappa_{ex}$ dominates. 

\subsubsection{Normalization of the heterodyne signal and the detection efficiency}
The heterodyne signal is normalized such that the variance of the vacuum signal is equal to 1. To this end, we define a normalized optical annihilation operator of the vacuum signal incident on the detector as
\begin{equation}
    A_{in}=\dfrac{1}{\sqrt{T}}\int_{0}^{T}~a_{in}(t)~dt.
\end{equation}
Here, $T$ is a finite detection time and $a_{in}(t)$ obeys the relations
\begin{eqnarray}
    &&\expval{a_{in}(t)a^{\dagger}_{in}(t')}=\delta(t-t')\label{seq:vacuum1}\\
    &&\expval{a_{in}^{\dagger}(t)a_{in}(t')}=0\label{seq:vacuum2}\\
    &&[a_{in}(t),a^{\dagger}_{in}(t')]=\delta(t-t'),\label{seq:vacuum3}
\end{eqnarray}
which describes vacuum statistics~[\href{https://www.google.co.uk/books/edition/Quantum_Noise/a_xsT8oGhdgC?hl=en&gbpv=1&pg=PA1&printsec=frontcover}{C.~W.~Gardiner and P.~Zoller, \emph{Quantum noise: a handbook of Markovian and non-Markovian quantum stochastic methods with applications to quantum optics} (Springer Science \& Business Media, 2004)}]. The variance of the optical quadrature $X_{in}=(A_{in}+A^{\dagger}_{in})/\sqrt{2}$ is then 
\begin{eqnarray}
 \mathrm{Var}\left(X_{in}\right)&=&\mathrm{Var}\left[\frac{1}{\sqrt{2}}\left(A_{in}+A^{\dagger}_{in}\right)\right]\nonumber\\
 &=&\frac{1}{2}\left[\expval{A^{\dagger}_{in}A_{in}+A_{in}A^{\dagger}_{in}}\right]\nonumber\\
 &=&\frac{1}{2}.
\end{eqnarray}
Moreover, the variance of all rotated optical quadratures $X_{in}(\theta)=(A_{in}e^{-i\theta}+A^{\dagger}_{in}e^{i\theta})/\sqrt{2}$ are equal to $1/2$ owing to the phase symmetry of the vacuum state. In contrast to homodyne detection, a heterodyne detector has extra vacuum noise owing to the simultaneous quadrature measurement. The heterodyne signal variance, multiplied by a normalization factor $\mathcal{N}(T)$, is
\begin{eqnarray}
    \mathrm{Var}\left(X_{het}\right)&=&\mathcal{N}(T)\left[\frac{1}{2}+\frac{1}{2}\right]\nonumber\\
    &=&\mathcal{N}(T).
\end{eqnarray}
Hence, to normalize the vacuum signal to 1, we divide the heterodyne variance measured over time $T$ by $\mathcal{N}(T)$.

The statistics of the optical mode will be modified by the dynamics of the cavity and the optomechanical interaction. The input-output relation for the anti-Stokes mode entering and exiting the cavity is $a_{out}(t)=\sqrt{2\kappa_{ex}}a(t)-a_{in}(t)$~[\href{https://doi.org/10.1103/PhysRevA.31.3761}{C.~W. Gardiner, and M.~J. Collett, Physical Review A \textbf{31}, 3761 (1985)}]. Integrating the input-output relation over the time $T$ then gives
\begin{eqnarray}
    A_{out}=\sqrt{2\kappa_{ex}}\dfrac{1}{\sqrt{T}}\int_{0}^{T}~a(t)~dt-A_{in},\label{seq:input-output1}
\end{eqnarray}
where
\begin{equation}
    A_{out}=\dfrac{1}{\sqrt{T}}\int_{0}^{T}~a_{out}(t)~dt.
\end{equation}
Once the steady state of laser cooling has been reached, the operator for the anti-Stokes cavity mode satisfies $\dot{a}=0$ and hence Eq.~\eqref{seq:input-output1} simplifies to
\begin{eqnarray}
    A_{out}=\sqrt{2\kappa_{ex}T}~a-A_{in}.\label{seq:input-output2}
\end{eqnarray}
Detection inefficiencies are then described by the beamsplitter transformation
\begin{eqnarray}
    A'_{out}=\sqrt{\eta}A_{out}+\sqrt{1-\eta}a_{v},
\end{eqnarray}
where $a_{v}$ is an annihilation operator of a normalized vacuum mode. The variance of the quadrature $X'_{out}=(A'_{out}+A^{'\dagger}_{out})/\sqrt{2}$ is then
\begin{eqnarray}
 \mathrm{Var}\left(X'_{out}\right)&=&\mathrm{Var}\left[\frac{1}{\sqrt{2}}\left(A'_{out}+A^{'\dagger}_{out}\right)\right]\nonumber\\
 &=&\frac{1}{2}\left[2\eta\kappa_{ex}T\expval{aa^{\dagger}+a^{\dagger}a}+\eta\expval{A_{in}A_{in}^{\dagger}+A_{in}^{\dagger}A_{in}}+(1-\eta)\expval{a_{v}a_{v}^{\dagger}+a_{v}^{\dagger}a_{v}}\right]\nonumber\\
 &=&\frac{1}{2}+2\eta\kappa_{ex} T\left(\expval{a^{\dagger}a}+\frac{1}{2}\right).
\end{eqnarray}
As before, the heterodyne adds extra vacuum noise and in the experiment, multiplies the signal by $\mathcal{N}(T)$ and hence
\begin{eqnarray}
    \mathrm{Var}\left(X_{het}\right)&=&\mathcal{N}(T)\left[\frac{1}{2}+ \mathrm{Var}\left(X'_{out}\right)\right]\nonumber\\
    &=&\mathcal{N}(T)\left[1+2\eta\kappa_{ex}T\left(\expval{a^{\dagger}a}+\frac{1}{2}\right)\right].
\end{eqnarray}

\subsubsection{Experimental parameters}

The parameters for our microresonator experiment are listed in Table~\ref{table:parameters}. Note that the optical and mechanical decay rates ($\kappa_{p}$/$\kappa$/$\gamma$) correspond to amplitude decay rates.

\begin{table}[ht]
\caption{Experimental parameters}
\centering 
\begin{tabular}{l c c c} 
\hline\hline 
\textbf{Parameter} & \textbf{Symbol} & \textbf{Value} & \textbf{Units} \\ [0.5ex] 
\hline 
Microsphere Diameter & $D$ & 280 & $\upmu$m \\
Pump Frequency & $\omega_{p}/2\pi$ & 193 & THz \\ 
Mechanical Frequency & $\omega_m/2\pi$ & 10.85 & GHz \\
Pump Power & & 372 & $\upmu$W \\ 
Pump Linewidth [FWHM] & $2 \kappa_{p} / 2 \pi$ & 6.1 & MHz \\ 
Pump External Linewidth [FWHM] & $2 \kappa_{p,ex} / 2 \pi$ & 2.9 & MHz \\ 
Anti-Stokes Linewidth [FWHM] & $2 \kappa / 2 \pi$ & 45.2 & MHz \\ 
Anti-Stokes External Linewidth [FWHM] & $2 \kappa_{ex} / 2 \pi$ & 10.1 & MHz \\
Mechanical Linewidth & $2 \gamma / 2 \pi$ & 45.0 & MHz \\
Optomechanical Coupling Rate & $g_{0} / 2 \pi$ & 300 & Hz \\
Pump Enhanced Optomechanical Coupling Rate & $G / 2 \pi$ & 3.3 & MHz \\
Detection Efficiency & $\eta$ & 0.32 & \% \\
Sample Temperature & & 295 & K \\
Initial Mean Photon Number & $\bar{N}$ & 558 & \\  
[1ex]
\hline \hline
\end{tabular}
\label{table:parameters}
\end{table}

\subsubsection{Resonators and tapers}

The resonator is manufactured in a Fujikura FSM100 fusion splicer. We place a cleaved optical fiber (SMF-28e+) into the splicer with a $2.2\,$mm arc gap and $3.9\,$s discharge. This forms a ($280\,\mu$m) silica sphere. This sphere is brought close to a tapered optical fiber. The tapered optical fiber is pulled using a tapering rig which comprises two stepper motors (MTS50A-Z8, Thorlabs) holding the fiber over a hydrogen flame regulated by a flow controller (El-flow base, Bronkhorst). We control the flame and tension applied by the stepper motors while monitoring the transmission of the fiber. Once the interference between higher order modes of the taper disappears (seen in fluctuating transmission) we stop the tapering process. This allows us to consistently pull a taper that operates with a single optical mode. The tapered fiber and resonator are kept in a flow box to prevent contamination from dust. 

\subsubsection{Determining cavity losses}

We are able to determine the intrinsic, $\kappa_{in}$, and external, $\kappa_{ex}$, cavity losses by adjusting the distance of the tapered fiber from the resonator and measuring the transmission spectrum at each position. Each transmission spectrum is fit to a Fano function of the form:
\begin{eqnarray}
    F(\omega)=A\left | e^{i \Delta\phi}-\frac{2\kappa_{ex}}{i(\omega-\omega_0) + (\kappa_{in} + \kappa_{ex})}\right |^{2} .
    \label{eq:fano}
\end{eqnarray}
\begin{figure}
    \centering
    \includegraphics[width=0.5\textwidth]{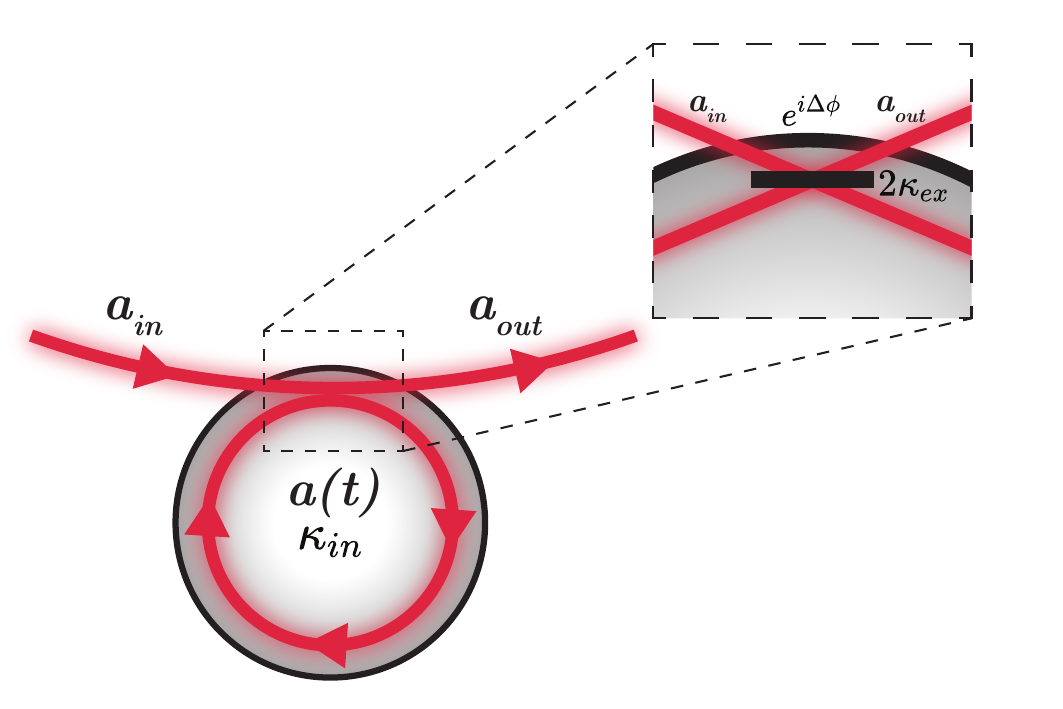}
    \caption{A simplified coupling configuration between an incoming waveguide mode $a_{in}$, a localised cavity mode $a$ and an outgoing waveguide mode $a_{out}$. Coupling is captured by a beamsplitter interaction between the incoming and cavity mode to produce the recirculated and outgoing fields. The phase difference $\Delta\phi$ between paths originates from the extended interaction region of a real taper coupling system -- it is included as a phenomenological parameter in this simplified model.}
    \label{fig:FanoCouplingConfig}
\end{figure}
This is derived by considering the coupling configuration displayed in Fig.~\ref{fig:FanoCouplingConfig} which uses a modified form of the input-output relation for a cavity in transmission
\begin{equation}
    a_{out}(t) = e^{i\Delta\phi}a_{in}(t) - \sqrt{2\kappa_{ex}}a(t),
    \label{eq:input-output_mod}
\end{equation}
where $a_{in}(t)$ ($a_{out}(t)$) is the travelling optical envelope annhiliation operator at the input (output) of the cavity, $\kappa_{ex}$ is the external amplitude coupling rate between the intra-cavity and extra-cavity fields and $\Delta\phi$ is a phase accumulated between the field travelling within the cavity and the field that traverses the waveguide. The equation of motion for the intra-cavity field is given by
\begin{equation}
    \Dot{a}(t) = -i\omega_0 a(t) - (\kappa_{in} + \kappa_{ex}) a(t) + \sqrt{2\kappa_{ex}}a_{in}(t) ,
    \label{eq:cavityEOM}
\end{equation}
where $\kappa_{in}$ is the intrinsic amplitude decay rate of the cavity and $\omega_0$ is the resonant frequency of the cavity. Taking the Fourier transform of the above expression, we obtain:
\begin{equation}
    -i\omega\Tilde{a}(\omega) = -i\omega_0 \Tilde{a}(\omega) - (\kappa_{in} + \kappa_{ex}) \Tilde{a}(\omega) + \sqrt{2\kappa_{ex}}\Tilde{a}_{in}(\omega) ,
\end{equation}
where $\Tilde{\cdot}$ represents a Fourier transformed quantity. Rearranging Eq.(\ref{eq:cavityEOM}) and substituting it into the Fourier transform of Eq.(\ref{eq:input-output_mod}), we obtain
\begin{equation}
    \Tilde{a}_{out} = \left(e^{i\Delta\phi} - \frac{2\kappa_{ex}}{i(\omega-\omega_0) + (\kappa_{in} + \kappa_{ex})} \right)\Tilde{a}_{in} .
\end{equation}
Finally, the transmission spectrum of the cavity $T(\omega)$ is given by
\begin{equation}
    T(\omega) = \abs{\frac{\Tilde{a}_{out}}{\Tilde{a}_{in}}}^2 = \abs{e^{i\Delta\phi} - \frac{2\kappa_{ex}}{i(\omega-\omega_0) + (\kappa_{in} + \kappa_{ex})}}^2 ,
\end{equation}
which is equal to Eq.(\ref{eq:fano}) up to a normalisation factor. Experimental spectra of the pump and anti-Stokes modes fitted using Eq.(\ref{eq:fano}) are shown in Fig.~\ref{fig:modekappas}.

\begin{figure}[t]
    \centering
    \includegraphics[width=\textwidth]{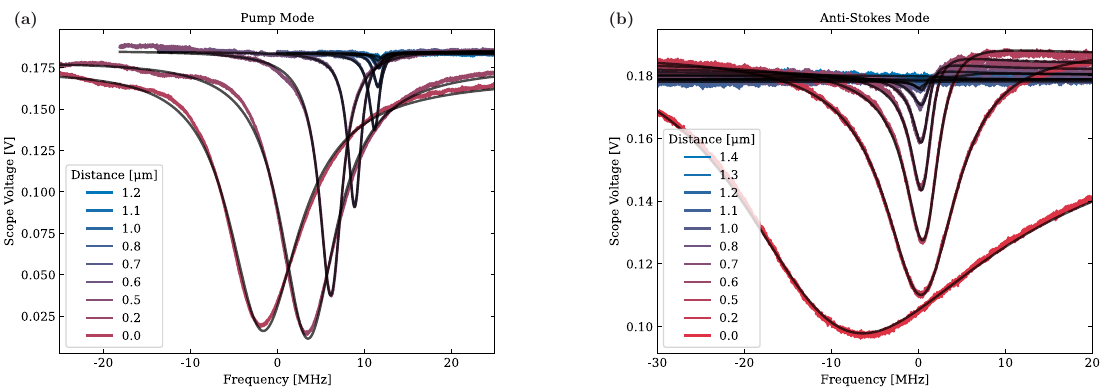}
    \caption{Transmission spectra for the optical pump mode (a) and the cavity anti-Stokes mode (b) for different taper-resonator distances. Coloured lines are experimental data and black lines are fits using Eq.~(\ref{eq:fano}).
}
    \label{fig:modekappas}
\end{figure}

\end{document}